\documentclass[smallextended]{svjour3}       

\usepackage[table,xcdraw,dvipsnames]{xcolor} 
\usepackage[utf8]{inputenc} 
\usepackage[T1]{fontenc}
\usepackage{graphicx}
\usepackage{glossaries}
\usepackage{soul} 
\usepackage[bookmarksdepth=2]{hyperref} 
\usepackage[misc,geometry]{ifsym}  
\usepackage{float} 
\usepackage[sort&compress,square,comma,numbers]{natbib}
\usepackage[many]{tcolorbox} 
\usepackage{enumitem} 
\usepackage{ifthen} 
\usepackage{amssymb} 
\usepackage{booktabs}
\usepackage{graphicx}
\usepackage{makecell}
\usepackage{amsmath}
\usepackage[export]{adjustbox}
\usepackage{ragged2e} 
\usepackage{relsize} 
\usepackage{multirow} 
\usepackage{pbox} 
\usepackage{balance} 
\usepackage{moresize} 
\usepackage{xspace}

\newtcolorbox{mybox}[2][]{
top=0.15in,left=4pt,right=4pt,bottom=4pt,
fonttitle=\bfseries,
colbacktitle=gray,
colback=gray!5,
colframe=gray!40!black,
enhanced,
attach boxed title to top left={xshift=1.5em,yshift=-\tcboxedtitleheight/2},
boxed title style={size=small},
drop shadow={black!50!white},
title=#2,#1}

\newcommand{\countobservations}{
    \def \countobservations{1}
}
\newcounter{observation}
\newenvironment{observation}[1]
{   \refstepcounter{observation}
    \noindent \textbf{Observation~\theobservation) \textit{#1}}
}

\countobservations

\newcommand{\countimplications}{
    \def \countimplications{1}
}
\newcounter{implication}
\countimplications

\newboolean{showcomments}
\setboolean{showcomments}{true}

\ifthenelse{\boolean{showcomments}}
{\newcommand{\nbc}[3]{
 {\colorbox{#3}{\bfseries\sffamily\scriptsize\textcolor{white}{#1}}}
 {\textcolor{#3}{\sf\small$\blacktriangleright$\textit{#2}$\blacktriangleleft$}}
 }
}
{\newcommand{\nbc}[3]{}
 }

\newcommand{\smalltt}[1]{\ifmmode{\mbox{\smaller\texttt{#1}}}\else{\smaller\tt #1}\fi}
\newcommand{\code}[1]{\smalltt{#1}}

\newcolumntype{L}[1]{>{\raggedright\let\newline\\\arraybackslash\hspace{0pt}}m{#1}}
\newcolumntype{C}[1]{>{\centering\let\newline\\\arraybackslash\hspace{0pt}}m{#1}}
\newcolumntype{R}[1]{>{\raggedleft\let\newline\\\arraybackslash\hspace{0pt}}m{#1}}

\usepackage{soul}
\usepackage{moresize}
\usepackage{listings}
\begin{document}

	\title{Predicting post-release defects with knowledge units (KUs) of programming languages: an empirical study}
	
	\titlerunning{KUs of PLs: \\A New Perspective for Studying the Source Code}
	
	\author{Md Ahasanuzzaman \and Gustavo A. Oliva \and Ahmed E. Hassan \and Zhen~Ming~(Jack)~Jiang}

	\institute{
		\Letter \space Md Ahasanuzzaman, Gustavo A. Oliva, and Ahmed E. Hassan \at
		Software Analysis and Intelligence Lab (SAIL), School of Computing \\
		Queen's University, Kingston, Ontario, Canada\\    
		\email{16ma87@queensu.ca,\{gustavo,ahmed\}@cs.queensu.ca} \\
        \Letter \space Zhen~Ming~(Jack)~Jiang \at York University, Toronto, Ontario, Canada \\
        \email{zmjiang@cse.yorku.ca} 
	}

	\date{Received: date / Accepted: date}

	\maketitle
    
	\begin{abstract}
		\justifying{Defect prediction plays a crucial role in software engineering, enabling developers to identify defect-prone code and improve software quality. While extensive research has focused on refining machine learning models for defect prediction, the exploration of new data sources for feature engineering remains limited. Defect prediction models primarily rely on traditional metrics such as product, process, and code ownership metrics, which, while effective, do not capture language-specific traits that may influence defect proneness. To address this gap, we introduce Knowledge Units (KUs) of programming languages as a novel feature set for analyzing software systems and defect prediction. A KU is a cohesive set of key capabilities that are offered by one or more building blocks of a given programming language. We conduct an empirical study leveraging 28 KUs that are derived from Java certification exams and compare their effectiveness against traditional metrics in predicting post-release defects across 28 releases of 8 well-maintained Java software systems. Our results show that KUs provide significant predictive power, achieving a median AUC of 0.82, outperforming individual group of traditional metric-based models (e.g., process, product and ownership metrics). Among KU features, Method \& Encapsulation, Inheritance, and Exception Handling emerge as the most influential predictors. Furthermore, combining KUs with traditional metrics enhances prediction performance, yielding a median AUC of 0.89. We also introduce a cost-effective model using only 10 features (5 KUs and 5 traditional metrics), which maintains strong predictive performance while reducing feature engineering costs. Our findings demonstrate the value of KUs in predicting post-release defects, offering a complementary perspective to traditional metrics. This study can be helpful to researchers who wish to analyze software systems from a perspective that is complementary to that of traditional metrics.}

		\keywords{Defect prediction , Software metrics, code metrics, Java, certification exam, software engineering, and machine learning}
	\end{abstract}

    \newcommand\RQOne{How effective are KU features for predicting post-release defects?}

    \newcommand\RQTwo{What are the most important KU features for predicting post-release defects?}

    \newcommand\RQThree{Can KU features provide insights into the incidence of post-release defects?}

    \newcommand\RQFour{Can KUM be made more accurate by combining its features with those from TM?}

    \newcommand\RQFive{How different classification algorithms with hyper-parameter tuning impact the combined model for predicting post-release defects?}

    \newcommand\RQSix{Can a cost-effective model be created by combining specific features from KUM and TM?}

    \newcommand\FUPOne{What knowledge units typically have the greatest predictive power?}
    \newcommand\FUPTwo{Which features are more decisive in the combined model? Traditional metrics or Knowledge units?}
        
    \newcommand\DiscussionOne{How well do models built with both traditional metrics and knowledge units perform in defect prediction?}
        
    \newcommand\DiscussionTwo{Can knowledge units provide insights into the incidence of post-release defects?}

    \section{Introduction}
\label{sec:Intro}

Defect prediction has been a long-standing area of research in the field of software engineering, aimed at identifying potential defect-prone code in software systems~\citep{koru2008investigation, nachi_metrics_component_failure, moser_change_static_code_metrics_defect_prediction, rahman2013and, majumder2022revisiting}. While the defect prediction field has seen substantial research in developing new machine learning approaches (e.g., bootstrapped by the recent advances and popularization of deep learning techniques~\citep{deepjit_defect_prediction, cnn_defect_prediction, chen2019deepcpdp, zain2023application}, the exploration of novel types of data for feature engineering remains less explored. The continual improvement in the defect prediction field demands innovation not just in modeling techniques but also in the data used for feature engineering. The more indicators researchers have, the better equipped they are to understand defects and hypothesize their underlying causes.


Product, process and code ownership metrics have been extensively used for defect prediction over several decades. We collectively refer to them as traditional metrics. These metrics characterize the size (e.g., LOC), complexity (e.g., cyclomatic complexity), and structure (e.g., the CK suite \citep{CK94}) of object-oriented software systems. Most importantly, these metrics are \textit{programming-language agnostic}, in the sense that their definitions are applicable to and computable for any object-oriented system (irrespective of the used programming language). As an inherent limitation, traditional metrics do not reveal system \textit{traits} that are tied to certain building blocks of a given programming language. As an illustrative example, let us consider a system written in Java and that makes use of the Java Concurrency API (i.e., a few classes from the \texttt{java.util.concurrent} package). Detecting the use of such an API reveals that concurrent code is a \textit{trait} of that system. Such a trait is relevant: specific defects might arise because of it, not every developer is proficient in concurrent programming, and debugging a concurrent program is often more difficult than debugging a single-threaded one. Yet, traditional metrics would not capture this concurrency trait. 


Towards filling this gap, this paper reports an empirical study on the usage of knowledge units (KUs) of the Java programming language for predicting post-release defects. As defined in our prior study~\citep{ahasanuzzaman2024using}, a KU is a \textit{cohesive set of key capabilities that are offered by one or more building blocks of a given programming language} (see Figure~\ref{fig:ku_metamodel}). The word ``capabilities'' refer to the main things that a programmer can do with a certain building block of a programming language. For instance, using the \textit{Concurrency API} building block, a developer can create worker threads to execute tasks concurrently. Therefore, it is reasonable to assume that Java has a \textit{Concurrency KU}, which includes a cohesive set of key concurrent program capabilities offered by the Concurrency API (building block). Our hypothesis is that by capturing system traits such as \textit{concurrency} through KUs, we will be able to design more accurate post-release defect prediction models compared to those relying exclusively on traditional metrics. 

We note that KUs are also able to capture more nuanced code characteristics (e.g., compared to concurrent programming) that may still influence defect-proneness. For instance, a \textit{method} is a fundamental building block in Java. Developers can create methods with parameters, overloaded methods, methods with variable-length arguments, and define access modifiers for methods. Not every object-oriented language supports these capabilities. Therefore, it is also reasonable to assume that Java has a \textit{Method \& Encapsulation KU}, which captures the key capabilities associated with methods and constructors. KUs capture deeper concepts than merely counting keywords. For instance, detecting the Method \& Encapsulation KU involves tracking the usage of variable length arguments in methods and the application of overloaded constructors and constructor chaining. To achieve this, we use the Eclipse JDT framework to parse the code's Abstract Syntax Tree (AST), allowing us to gather nuanced and richer information that goes much beyond simple keyword matching. By incorporating such granular details, KUs offer a more comprehensive perspective on software structure that extends beyond traditional metrics. 

As described in our prior work~\citep{ahasanuzzaman2024using}, we elicit KUs by manually analyzing three certification exams for the Java programming language (Section~\ref{subsec:knowledge_operational_definition}): Java SE 8 Programmer I Exam, Java SE 8 Programmer II Exam, and Java EE Developer Exam. We elicit a total of 28 KUs, which cover a wide range of capabilities (\ref{appendix:cert-exams}). To count the occurrences of each KU in a Java source code file, we built a custom KU detector on top of the Eclipse JDT framework (Section~\ref{subsec:knowledge_detection}). We refer to the set of metrics corresponding to the counts of each KU as \textit{KU metrics}.

Our data collection is designed as follows. First, we select the defect dataset that was curated by~\citet{Yatish_post_release_defect}. This dataset contains post-release defect data for 28 releases that originate from 8 Java systems. For each source code file in a release, we compute our 28 KU metrics. Additionally, we reuse all 65 traditional metrics provided in the defect dataset, including 54 product metrics, 5 process metrics, and 6 code ownership metrics for each source code file. Finally, we build defect prediction models using these two sets of metrics to address the following research questions (RQs):



\smallskip \noindent \textbf{\textbf{RQ1:} {\RQOne}} First, we use the KU metrics to build a defect prediction model for each of the studied releases. We refer to this model as KUM. To judge the performance of KUM, we build four baseline models using traditional metrics: (i) PROD (which is built with product metrics), (ii) PROC (which is built with process metrics), (iii) OWN (which is built with code ownership metrics) and (iv) TM (which is built using all product, process and code ownership metrics). To compare the performance of KUM and any of the baseline models, we perform a statistical test using a two-sided Wilcoxon signed-rank test. Additionally, we employ Cliff's delta~\citep{cliff_delta} as an effect size measure to assess the practical significance of the differences between the distributions. We also rank the studied models using the Scott-Knott ESD (SK-ESD) technique~\citep{ghotra_feature_importance_ICSME, mittas_ranking_feature_TSE, kla_model_validation}. Our results indicate that:

\sloppy\smallskip\noindent\textit{KUM achieves a median AUC of 0.82 (a defect prediction model with an AUC of 0.70 or higher is typically considered suitable for practical use). KUM outperforms three baseline models: PROD, PROC and OWN. The normalized AUC improvement of KUM over PROD ranges from 3.6\% to 22.2\% (median 11.4\%), improvement over PROC ranges from 6.7\% to 68.9\% (median 42.4\%), and improvement over OWN ranges from 3.3\% to 71.9\% (median (33.3\%). That is, KUM not only outperforms a model built on the same raw data (PROD), but also those built on process (PROC) and ownership (OWN) data. TM is the top-performing model, achieving a median AUC of 0.85}

\smallskip \noindent \textbf{\textbf{RQ2:} {\RQTwo}} To determine the most important features of KUM, we utilize SHAP (SHapley Additive exPlanation). SHAP is a robust, flexible, and widely used method for model interpretation~\citep{molnar2020interpretable,Gopi22}. Next, we apply the SK-ESD technique on the distribution of SHAP values to rank features. Our results indicate that:

\smallskip \noindent \textit{Method \& Encapsulation, Inheritance, and Exception are identified as the top three most important KU features of KUM for predicting post-release defects.}



\smallskip \noindent \textbf{\textbf{RQ3:} {\RQFour}} Based on the encouraging results from RQ1, we join all the features of KUM and TM to build a combined model, which we call KUM+TM. We compare the performance of KUM+TM with that of KUM and TM using the same approach from RQ1. To investigate the importance of feature dimensions in KUM+TM, we employ the same SHAP approach from RQ2. Our results indicate that:

\smallskip\noindent\textit{KUM+TM achieves a median AUC of 0.89 and outperforms both KUM and TM across all of the studied releases. The normalized AUC improvement of KUM+TM over TM ranges from 11.1\% to 44.4\% (median 22.5\%) and the improvement over KUM ranges from 10.5\% to 72.2\% (median 31.5\%) across the studied releases. The most important feature is the number of active developers (ADEV, a process feature). The top ranked KU feature is \textit{Method \& Encapsulation} at rank 4. \textit{Inheritance} and \textit{Exception Handling} KU features are also among the top ten most important features.}



\smallskip \noindent \textbf{\textbf{RQ4:} {\RQSix}} 
A cost-effective model that uses fewer features and yet maintains decent performance is valuable, as it reduces operational costs associated with extensive feature engineering. Towards this goal, we build a cost-effective combined model, which we call COST\_EFF. This model is built using the top five features of KUM and the top five traditional metric features of TM. This results in a combined model with only 10 features, compared to the 65 features in TM and the 28 features in KULTC. Our results indicate that:

\sloppy\smallskip\noindent\textit{The COST\_EFF achieves a median AUC of 0.87 and outperforms TM. The normalized AUC improvement of COST\_EFF over TM ranges from 3.7\% to 31.3\% (median 11.1\%). Therefore, we improve over the model built with all traditional metrics while still keeping feature engineering cost at a much more reasonable level.}

The main contribution of this paper is to evaluate the effectiveness of KUs of the Java programming language for predicting post-release defects. We believe that our findings can be helpful and encouraging to researchers who wish to analyze software systems from a perspective that is complementary to traditional metrics. Our encouraging results indicate that further refining the concept of KUs and exploring alternative elicitation techniques can potentially lead to even higher-performing defect prediction models. To bootstrap future work in this area, we provide online supplementary material with the data analyzed as part of this study\footnote{\url{http://bit.ly/3GoSmHL}}.

\smallskip \noindent \textbf{Paper organization.} Section~\ref{sec:knowledge_unit} defines KUs and presents our approach for detecting them. Section~\ref{sec:Data_Collection} describes our data collection approach. Section~\ref{sec:Findings} presents the motivation, approach, and findings that are associated with our research questions. Section~\ref{sec:discussion} discusses the trade-offs between traditional metrics and KUs for defect prediction and future directions to expanding KUs. Section~\ref{sec:Limitations_And_Threats} describes the threats to the validity of our findings. Section~\ref{sec:Related_Work} discusses related work. Finally, Section~\ref{sec:Conclusion} concludes the paper.
    \section{Knowledge Units (KUs)}
\label{sec:knowledge_unit}

We introduce the concept of knowledge units (KUs) of programming language in our prior work~\citep{ahasanuzzaman2024using}. To make this paper as self-contained as possible, in the following we provide a brief introduction to KUs (Section~\ref{subsec:knowledge_definition}) and discuss our approaches for eliciting (Section~\ref{subsec:knowledge_operational_definition}) and detecting them (Section~\ref{subsec:knowledge_detection}).

\subsection{Definition}
\label{subsec:knowledge_definition}

\begin{figure}[!t]
	\centering
    \includegraphics[width=1.0\linewidth]{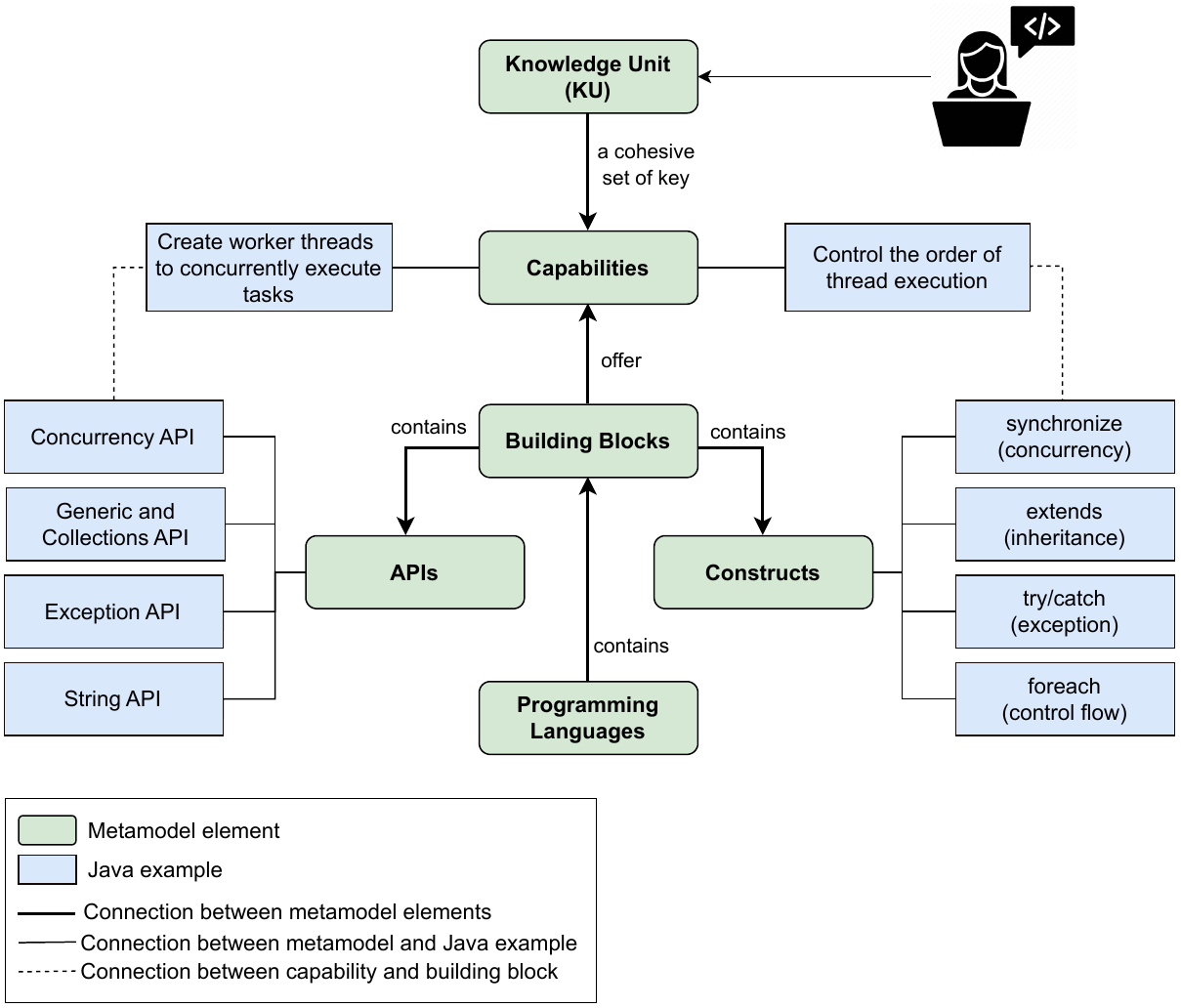} 
    \caption{Our metamodel for knowledge units (KUs).}
	\label{fig:ku_metamodel}
\end{figure}

Every programming language has its own building blocks. The building blocks of a programming language are the ``blocks'' that developers use to ``build'' (write) code. In the case of Java, we consider them to be fundamental language constructs (e.g., arrays, if/else statements, try/catch statements) and APIs (e.g., Generic and Collection API, Concurrency API and String API). Each building block offers a \textit{set of capabilities} essentially representing the actions a developer can perform using that building block. We define a Knowledge Unit (KU) as a cohesive set of \textbf{key} capabilities that are offered by \textbf{one or more building blocks} of a given programming language. The inclusion of ``key'' in our definition aims to ensure that KUs are centered around fundamental capabilities rather than those that are overly specific. Figure~\ref{fig:ku_metamodel} illustrates key capabilities that are associated with the \texttt{Concurrency API} (left-hand side) and the \texttt{synchronize} language construct (right-hand side). 


\subsection{Eliciting KUs}
\label{subsec:knowledge_operational_definition}

Certification exams of a programming language (e.g., Oracle Java SE and Java EE certification exams for Java) aim to determine the proficiency of a developer in using the key capabilities offered by the building blocks of that language. Therefore, we can  say that \textit{certification exams capture the KUs of a programming language}. For instance, one of the topics covered in the Java SE 8 Programmer II certification exam is ``Generics and Collections.'' The subtopics of  ``Generics and Collections'' that are covered in the exam include: (i) create and use a generic class, (ii) create and use \texttt{ArrayList}, \texttt{TreeSet}, \texttt{TreeMap}, and \texttt{ArrayDeque} objects, (iii) use \texttt{java.util.Comparator} and \texttt{java.lang.Comparable} interfaces, and (iv) iterate using \texttt{forEach} methods of a \texttt{List}. We interpret such a list of subtopics as the key capabilities that are offered by the \textit{Generics and Collections} building block of the Java programming language. From this interpretation, we infer that Java has a \textit{Generics and Collections} KU.

We reuse the KUs that we elicited in our prior work~\citep{ahasanuzzaman2024using}. To elicit those KUs, we relied on the Oracle certification exams for the Java programming language. Oracle offers certification exams for different Java editions, such as Java Standard Edition (Java SE) and Java Enterprise Edition (Java EE). Specifically, for Java SE, there are two levels of certification exams:(i) \textit{Java SE 8 Programmer I Certification exam}~\citep{oracle_se_oap}, and (ii) \textit{Java SE 8 Programmer II Certification exam}\citep{oracle_se_ocp}. For Java EE, only one version of the exam is offered by Oracle: \textit{Oracle Certified Professional, Java EE Application Developer Certification exam}~\citep{oracle_ee_prof}. We elicited KUs directly from the topics outlined in these exams. In most cases, an exam topic was interpreted as a KU (e.g., the Generics and Collections topic of Java SE 8 Programmer II Certification exam was interpreted as ``Generics and Collections'') and its subtopics (e.g., create a generic class, use \texttt{TreeSet, TreeMap and ArrayList} and use \texttt{java.util.Comparator}) were interpreted as the key capabilities of that KU. At the end of this process, we elicited 28 KUs (Table~\ref{tab:topic_definition}). We refer to our prior work~\citep{ahasanuzzaman2024using} for the detailed description of this mapping process.




\begin{table}[!htbp]
    \centering
    \caption{The list of knowledge units (KUs) elicited for the Java programming language.}
    \label{tab:topic_definition}
    \resizebox{\textwidth}{!}{
    \begin{tabular}{p{4cm}p{12cm}}
        \toprule
        \multicolumn{1}{C{4cm}}{\textbf{Knowledge unit (KU)}} & \multicolumn{1}{C{12cm}}{\textbf{Definition}}                                                                                                                                                                                                                               \\ \midrule

        \textbf{[K1]} Data Type                                         &
        The declaration and initialization of different types of variables (e.g., primitive type and parameterized type.)
        \\ \midrule

        \textbf{[K2]} Operator \& Decision                             &
        The usage of different Java operators (e.g., assignment, logical, and bit-wise operators) and conditional statements (e.g, if, if-else, and switch statements).
        \\ \midrule

        \textbf{[K3]} Array                                            &
        The declaration, instantiation, initialization and the usage of one-dimensional and multi-dimensional arrays.
        \\ \midrule

        \textbf{[K4]} Loop                                             &
        The execution of a set of instruction-s/methods repeatedly using for, while, and do-while statements and the skipping and stopping of a repetitive execution of instructions and methods using continue and break statements.
        \\ \midrule

        \textbf{[K5]} Method \& Encapsulation                         &
        The creation of methods with parameters, the use of overloaded methods and constructors, the usage of constructor chaining, the creation of methods with variable length arguments, and the usage of different access modifiers. This KU also describes encapsulation mechanisms, such as creating a set and get method for controlling data access, generating immutable classes, and updating object type parameters of a method.
        \\ \midrule

        \textbf{[K6]} Inheritance                                      &
        Developing code with child class and parent class relationship, using polymorphism (e.g., developing code with overridden methods), creating abstract classes and interfaces, and accessing methods and fields of the parent's class.
        \\ \midrule

        \textbf{[K7]} Advanced Class Design                             &
        Developing code that uses the final keyword, creating inner classes including static inner classes, local classes, nested classes, and anonymous inner classes, using enumerated types including methods and constructors in an enum type, and developing code using @override annotator.
        \\ \midrule

        \textbf{[K8]} Generics \& Collection                           &
        The creation and usage of generic classes, usage of different types of interfaces (e.g., List Interface, Deque Interface, Map Interface, and Set Interface), and comparison of objects using interfaces (e.g., java.util.Comparator, and java.lang.Comparable).
        \\ \midrule

        \textbf{[K9]} Functional Interface                         &
        The development of code that uses different versions of defined functional interfaces (e.g., primitive, binary, and unary) and user-defined functional interfaces.
        \\ \midrule

        \textbf{[K10]} Stream API                                       &
        The development of code with lambda expressions and Stream APIs. This includes developing code to extract data from an object using peek() and map() methods, searching for data with search methods (e.g., findFirst) of the Stream classes, sorting a collection using Stream API, iterating code with foreach of Stream, and saving results to a collection using the collect method.
        \\ \midrule

        \textbf{[K11]} Exception                                      &
        The creation of try-catch blocks, the usage of multiple catch blocks, the usage of try-with-resources statements, the invocation of methods throwing an exception, and the use of assertion for testing invariants.
        \\ \midrule

        \textbf{[K12]} Date time API                                   & This KU refers to create and manage date-based and time-based events using Instant, Period, Duration, and TemporalUnit, and work with dates and times across timezones and manage changes resulting from daylight savings, including Format date and times values. \\ \midrule

        \textbf{[K13]} IO                                              &
        Reading and writing data from console and files and using basic Java input-output packages (e.g., java.io.package).
        \\ \midrule

        \textbf{[K14]} NIO                                              &
        Interacting files and directories with the new non-blocking input/output API (e.g., using the Path interface to operate on file and directory paths). Performing other file-related operations (e.g., read, delete, copy, move, and managing metadata of a file or directory).
        \\ \midrule

        \textbf{[K15]} String Processing                                &
        The knowledge about searching, parsing, replacing strings using regular expressions, and using string formatting.
        \\ \midrule

        \textbf{[K16]} Concurrency                                      &
        The knowledge about functionalities that are related to thread execution and parallel programming. Some of these functionalities include: creating worker threads using Runnable and Callable classes, using an ExecutorService to concurrently execute tasks, using synchronized keyword and java.util.concurrent.atomic package to control the order of thread execution, using java.util.concurrent  classes and using a parallel fork/join framework.
        \\ \midrule

        \textbf{[K17]} Database                                        &
        The creation of database connection, submitting queries and reading results using the core JDBC API.
        \\ \midrule   

        \textbf{[K18]} Localization                                     &
        The knowledge about reading and setting the locale (Oracle defines a locale as ``a specific geographical, political, or cultural region'') by using the Locale object, and building a resource bundle for each locale, and loading a resource bundle in an application.
        \\ \midrule

        \textbf{[K19]} Java Persistence                                  &
        The usage of object/relational mapping facilities for managing relational data in Java applications. With this knowledge, developers can learn how to map, store, update and retrieve data from relational databases to Java objects and vice versa.                                                              \\ \midrule

        \textbf{[K20]} Enterprise Java Bean                             &
        The knowledge about managing server-side components that encapsulate the business logic of an application.                                                                                                                                                                                                               \\ \midrule

        \textbf{[K21]} Java Message Service API                         &
        The knowledge of how to create, send, receive and read messages using reliable, asynchronous, and loosely coupled communication.                                                                                                                                                                                                   \\ \midrule

        \textbf{[K22]} SOAP Web Service                               &
        Creating and using the Simple Object Access Protocol for sending and receiving requests and responses across the Internet using JAX-WS and JAXB APIs.                                                                                                                                                          \\ \midrule

        \textbf{[K23]} Servlet                                          &
        Handling HTTP requests, parameters, and cookies and how to process them on the server sites with appropriate responses.                                                                                                                                                                                           \\ \midrule

        \textbf{[K24]} Java REST API                                    &
        Creating web services and clients according to the Representational State Transfer architectural pattern using JAX-RS APIs.                                                                                                                                                                                     \\ \midrule

        \textbf{[K25]} Websocket                                         &
        Creating and handling bi-directional, full-duplex, and real-time communication between the server and the web browser.                                                                                                                                                                                         \\ \midrule

        \textbf{[K26]} Java Server Faces                                 &
        The knowledge about how to build UI component-based and event-oriented web interfaces using the standard JavaServer Faces (JSF) APIs.                                                                                                                                                                                                     \\ \midrule

        \textbf{[K27]} Contexts and Dependency Injection (CDI)             &
        Managing the lifecycle of stateful components using domain-specific lifecycle contexts and type-safely inject components (services) into client objects.                                                                                                                                                     \\ \midrule

        \textbf{[K28]} Batch Processing                                    &
        Creating and managing long-running jobs on schedule or on demand for performing on bulk-data, and without manual intervention.                                                                                                                                                                                 \\ \bottomrule
    \end{tabular}
    }
\end{table}

\subsection{Detection of KUs}
\label{subsec:knowledge_detection}

Analogously to our prior work~\citep{ahasanuzzaman2024using}, we employ static analysis to detect KUs from a given Java project's release. First, we parse all Java source files of the release using the Eclipse JDT framework \citep{eclipse_jdt}. The JDT framework creates an Abstract Syntax Tree (AST) and provides visitors for every element of such tree, including \textit{names}, \textit{types} (class, interfaces), \textit{expressions}, \textit{statements}, and \textit{declarations}.\footnote{The grammar employed by JDT can be seen at \url{https://github.com/eclipse-jdt/eclipse.jdt.core/blob/master/org.eclipse.jdt.core.compiler.batch/grammar/java.g}}

To detect a KU, we resort to the set of key capabilities of that unit (Tables~\ref{tab:ku-from-java-exams}). For instance, to detect the presence of the \textit{Generics and Collections KU}, we use the visitors provided by JDT to detect the capabilities associated with such unit, namely: (i) create and use a generic class, (ii) create and use \code{ArrayList}, \code{TreeSet}, \code{TreeMap}, and \code{ArrayDeque} objects, and (iii) use \code{java.util.Comparator} and \code{java.lang.Comparable} interfaces. For instance, using type visitors, we can determine whether a given class is a generic class. Similarly, using statement visitors, we can find instantiations of an ArrayList. We also use JDT to collect binding information of methods, classes or variables (e.g., \code{org.eclipse.jdt.core.dom.IVariableBinding} resolves the binding information for a variable). In our study, we exclude type bindings that are related to third-party libraries since our primary interest lies in studying the KUs that are associated with the source code written by the developers of the studied projects. 

We count the occurrences of KUs for every Java source file. For example, one of the capabilities associated with the \textit{Generics and Collections KU} is to create a generic class. Hence, if we find a declaration of a \texttt{generic} class in a given Java file \code{X.java} belonging to a release \textit{R}, we increment the counter for $\textit{Generic and Collection KU}$ that is associated with file \texttt{X.java} by one. 
    \section{Data Collection}
\label{sec:Data_Collection}

\begin{figure}[!t]
	\centering
    \includegraphics[width=1.0\textwidth]{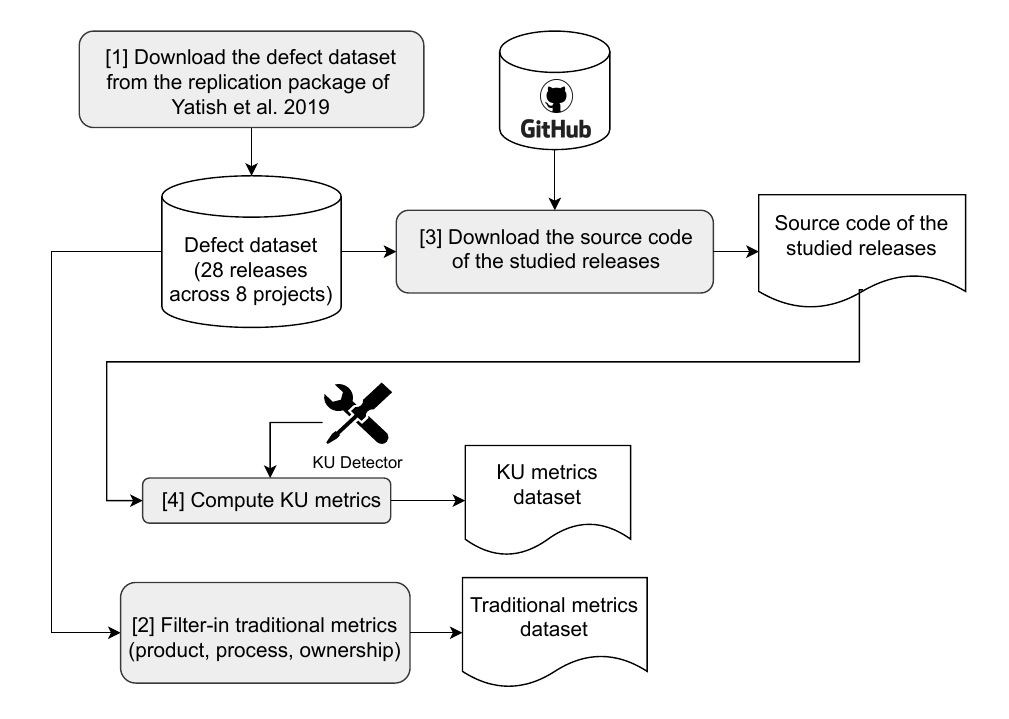} 
    \caption{An overview of our data collection process.}
	\label{fig:data_collection}
\end{figure}

Our data collection process is straightforward and contains four steps (Figure~\ref{fig:data_collection}). We collect the defect dataset from the replication package of~\citet{Yatish_post_release_defect} (Step 1) and filter-in traditional metrics (Step 2). Then, we download the source code of our studied releases (Step 3) and compute the per-file KU metrics using our custom KU detector (Step 4). In the following, we describe each step of our data collection process in more detail.

\begin{itemize}[label = \textbullet, wide = 0pt]
    \item \textbf{(Step 1) Download the defect dataset.} We choose the defect dataset that was curated by~\citet{Yatish_post_release_defect}. Instead of relying on heuristics to identify post-release defects, the authors employ an approach that ``leverages the earliest affected release that is realistically estimated by a software development team for a given defect''. We download the defect dataset from the article's replication package\footnote{\url{https://github.com/awsm-research/replication-icse2019}}. The dataset contains post-release defect data for 28 releases that span a total of 8 Java project. Table~\ref{tab:stat_defect_dataset} presents a statistical summary of the dataset (adapted from original paper).

    \begin{table}[!tbh]
        \caption{A statistical summary of the dataset.}
        \label{tab:stat_defect_dataset}
        \resizebox{\textwidth}{!}{
        \begin{tabular}{@{}llrrrr@{}}
        \toprule
        \multicolumn{1}{c}{\textbf{\begin{tabular}[c]{@{}c@{}}Project\\ Name\end{tabular}}} & \multicolumn{1}{c}{\textbf{\begin{tabular}[c]{@{}c@{}}Studied\\ Releases\end{tabular}}} & \multicolumn{1}{c}{\textbf{\begin{tabular}[c]{@{}c@{}}Number of\\ Java Files\end{tabular}}} & \multicolumn{1}{c}{\textbf{\begin{tabular}[c]{@{}c@{}}Line of \\ Code\end{tabular}}} & \multicolumn{1}{c}{\textbf{\begin{tabular}[c]{@{}c@{}}Total Defects\\ Reported\end{tabular}}} & \multicolumn{1}{c}{\textbf{\begin{tabular}[c]{@{}c@{}}Rate of Defects\\ (across releases)\end{tabular}}} \\ \midrule
        ActiveMQ                          & 5.0.0, 5.1.0, 5.2.0, 5.3.0                                                              & 1,884-3,420                                                                                 & 142-299K                                                                             & 3,157                                                                                         & 6\%-15\%                                                                                                 \\
        Derby                             & 10.2.1.6, 10.3.1.4, 10.5.1.1                                                            & 1,963-2,705                                                                                 & 412-533K                                                                             & 3,731                                                                                         & 14\%-33\%                                                                                                \\
        Groovy                            & 1.5.7, 1.6.0.Beta\_1, 1.6.0.Beta\_2                                                     & 755-884                                                                                     & 74-90K                                                                               & 3,943                                                                                         & 3\%-8\%                                                                                                  \\
        HBase                             & 0.94.0, 0.95.0, 0.95.2                                                                  & 1,059-1,834                                                                                 & 246-534K                                                                             & 5,360                                                                                         & 20\%-26\%                                                                                                \\
        Hive                              & 0.9.0, 0.10.0, 0.12.0                                                                   & 1,416-2,662                                                                                 & 287-563K                                                                             & 3,306                                                                                         & 8\%-19\%                                                                                                 \\
        JRuby                             & 1.1, 1.4.0, 1.5.0, 1.7.0                                                                & 731-1,614                                                                                   & 105-238K                                                                             & 5,475                                                                                         & 5\%-18\%                                                                                                 \\
        Lucene                            & 2.3.0, 2.9.0, 3.0.0, 3.1.0                                                              & 8,052-2,806                                                                                 & 101-242K                                                                             & 2,316                                                                                         & 3\%-24\%                                                                                                 \\
        Wicket                            & 1.3.0.beta1, 1.3.0beta2, 1.5.3                                                          & 1,672-2,578                                                                                 & 109-165K                                                                             & 3,327                                                                                         & 4\%-7\%                                                                                                  \\ \bottomrule
        \end{tabular}
        }
        \end{table}

    \item \textbf{(Step 2) Filter-in traditional metrics (product, process, and ownership)} We filter-in the traditional metrics from the defect dataset. This filtered dataset contains 54 product, 5 process, and 6 code ownership metrics (65 total) calculated for each Java source code file. Product metrics were calculated using the Understand tool~\citep{undertand_tool}. Examples include the number of lines of code (CountLineCode), number of declared public methods in a class (CountDeclMethodPublic), and number of other classes to which a class is coupled to (CountClassCoupled). The process metrics are: the number of commits (COMM), the number of lines added (ADDED\_LINES), the number of lines deleted (DEL\_LINES), the number of active developers (ADDEV) and the number of distinct developers (DDEV). The ownership metrics are: number of minor authors (developers contributing less than 5\% of the code in a file), number of major authors (developers contributing 5\% or more of the code in a module), and ownership ratio (the proportion of lines of code written by the developer who has the highest contribution of lines of code on the module). These ownership metrics are calculated for two granularities of code changes: lines of code (MINOR\_LINE, MAJOR\_LINE, and OWN\_LINE) and commit (MINOR\_COMMIT, MAJOR\_COMMIT, and OWN\_COMMIT). 
    
    We refer to these 65 metrics as \textit{traditional} because they have frequently been used in the literature to build and analyze defect prediction models~\citep{cross_project_kmeans,Varela17, jiarpakdee2020featureselection, laaber2021predicting, esteves2020understanding}. The full list of the traditional metrics can be seen in~\ref{appendix:code-metrics}.
    
    \item \textbf{(Step 3) Download the source code of the studied releases.} To download the source code of the studied software projects, we clone the repository of each project using the command git clone \texttt{project\_url} where \texttt{project\_url} is the url of the repository of the project in GitHub. Then, we extract the source code of a particular release using the command \texttt{git checkout tags/tag\_name}. Here, \texttt{tag\_name} is the name of one of the studied release for the project.

    \item \textbf{(Step 4) Compute KU metrics.} We count the occurrences of each KU in each Java file from every studied release in the defect dataset using our custom KU detector (refer to Section~\ref{subsec:knowledge_detection} for details).
\end{itemize}

The outputs of the data collection process are the \textit{KU metrics dataset} and the \textit{traditional metrics dataset} (Figure~\ref{fig:data_collection}). These datasets are employed to build several defect prediction models as part of our RQs.

    \section{Findings}
\label{sec:Findings}

In this section, we address our research questions. For each question, we discuss our motivation for studying it, the approach that we used to answer it, and our findings.

\subsection{RQ1: \RQOne}
\label{sec:section_rq1}

\smallskip \noindent \textbf{Motivation.} Traditionally, defect prediction has relied heavily on process, product, and code ownership metrics~\citep{gyimothy_TSE_metrics_fault_prediction,moser_change_static_code_metrics_defect_prediction,zhang_relationship_line_code_defects,emad_code_process_metrics} (traditional metrics). KUs offer a new lens through which the code of software systems can be analyzed, since they capture system traits that are not captured by those traditional metrics. More generally, we believe that exploring novel types of data for feature engineering is as important as developing or assessing new machine learning approaches for defect prediction. The more indicators researchers have, the better equipped they are to understand defects and hypotheisze their underlying causes.

\smallskip \noindent \textbf{Approach.} First, we build a defect prediction model using the KU metrics dataset for each of the studied releases. We treat each KU metric as a model feature and we refer to the entire set of KU features as the \textit{KU feature set}. Also, we refer to this KU model as KUM . Next, we evaluate the performance of KUM  based on an \textit{out-of-sample bootstrap} model validation technique. Finally, we compare the performance of KUM  to baseline models that are built using the traditional metrics dataset. In the following, we explain our approach in more detail:

\begin{figure}[!t]
	\centering
    \includegraphics[width=1.0\textwidth]{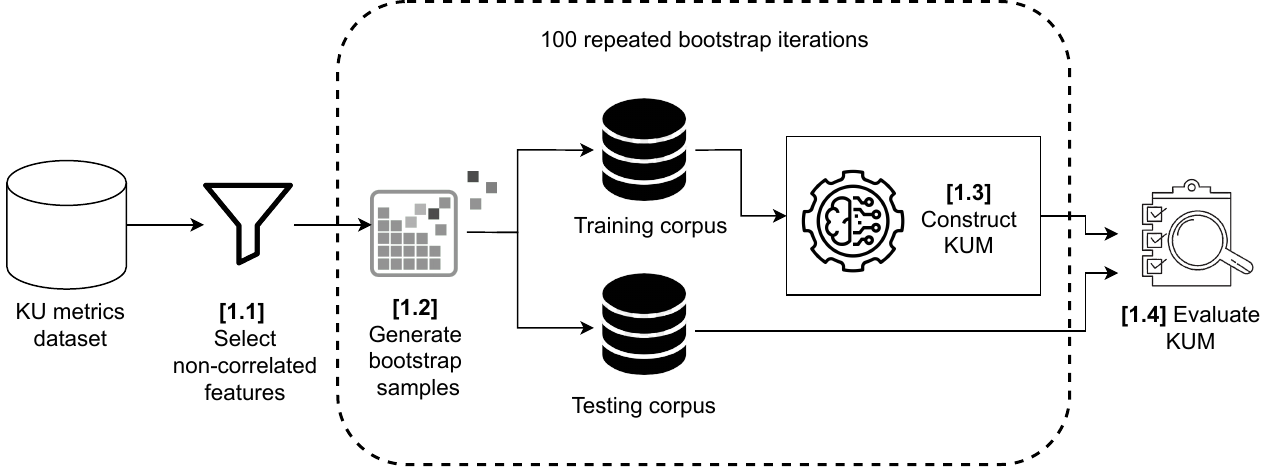} 
    \caption{An overview of our approach for building and evaluating KUM.}
	\label{fig:kucls_model_building}
\end{figure}

\begin{itemize}[itemsep=3pt, wide = 0pt, topsep=1pt, listparindent=\parindent]
    
    \item \textit{(Step 1) Construct KUM.} In this step, we construct and evaluate KUM. KUM is a binary classification model that outputs either \textit{clean} or \textit{defective} for an input Java file. An overview of our approach for building KUM can be seen in Figure~\ref{fig:kucls_model_building}.

    \begin{itemize}[itemsep=3pt, wide = 0pt, topsep=1pt, listparindent=\parindent]
        \item \textit{(Step 1.1) Select non-correlated features.} Prior to building any defect prediction model, it is recommended to mitigate correlated features~\citep{jiarpakdee2019impact} to better interpret the results of a model. Therefore, we want to check the performance of our studied models using non-correlated features. We select non-correlated features using AutoSpearman~\citep{jiarpakdee2018autospearman}. AutoSpearman automatically selects non-correlated features based on two analyses: (1) a Spearman rank correlation test and (2) a Variance Inflation Factor (VIF) analysis. Applying AutoSpearman\footnote{https://xai4se.github.io/defect-prediction/data-preprocessing.html} to the KU feature set resulted in the selection of all features, as no feature correlations were detected.

        \item \textit{(Step 1.2) Generate bootstrap samples.} To estimate the performance of our models in practice, we employ the \textit{out-of-sample bootstrap} validation technique with 100 repetitions~\citep{kla_model_validation, efron1983estimating}. In this technique, we randomly sample dataset with replacement to create 100 bootstrap samples. Each bootstrap sample has the same number of data points as the original dataset, but some points may appear multiple times while others may be excluded. These excluded data points from each bootstrap sample form the corresponding \textit{out-of-sample data}, which we use for testing our models. We train our models on each bootstrap sample and test its performance on the associated out-of-sample data.
        
        \item \textit{(Step 1.3) Construct KUM.} We use the KU feature set to construct KUM . Similarly to~\citet{Yatish_post_release_defect}, we use a random forest classifier with default parameters values (\texttt{scikit-learn} python implementation). For a detailed analysis of the influence of classifier choice and hyperparameter tuning on model performance, please refer to~\ref{appendix:hyperparameter}.

        \item \textit{(Step 1.4) Evaluate KUM.} To evaluate the accuracy of our defect prediction models, we calculate the \textit{Area Under the receiver operator characteristic Curve} (AUC). AUC is a robust, threshold-independent measure \citep{Chak_ICSE_SEIP}. AUC value ranges from 0 to 1, with 0.5 being as good as random guessing. The AUC values closer to one indicate a model that is good at distinguishing the class of interest (defective) from the other class (clean). Values closer to zero indicate a model that is performing worse than random guessing, consistently misclassifying instances by predicting the opposite class (i.e., labeling defective instances as clean and vice versa). Since we are employing an \textit{out-of-sample bootstrap} model validation with 100 repetitions, we obtain 100 AUC values for each prediction model that we evaluate.  
    \end{itemize}

    \item \textit{(Step 2) Construct baseline models.} Our main goal in RQ1 is to understand the efficacy of KUs in predicting software defects. To construct baseline models, we use the traditional metrics dataset that we extracted from our data collection. We construct four baseline model:  (i) a model built with product metrics (PROD), (ii) a model built with process metrics (PROC), (iii) a model built with code ownership metrics (OWN) and (iv) a model built with product, process, and ownership metrics (i.e., all the traditional metrics -- TM). These baseline models are constructed following the same procedure described in Step 1.
    
    \item \textit{(Step-3) Compare models' performance.} To check whether the performance difference between a given pair of models (e.g., KUM  vs PROD) is statistically significant, we apply the Wilcoxon signed-rank test. The Wilcoxon signed-rank test is a non-parametric statistical test for paired data. We infer that there is a statistically significant difference between the two input distributions when the p-value computed by the test is smaller than 0.05 (i.e., $alpha = 0.05$). In addition, we also use the Cliff's delta ($d$) effect size measure to quantify the practical difference between the two distributions \citep{cliff_delta}. We use the following thresholds for interpreting $d$ \citep{Romano06}: \textit{negligible} for $|\delta| \le 0.147$, \textit{small} for $0.147 < |\delta| \leq 0.33$, \textit{medium} for $0.33 < |\delta| \leq 0.474$, and \textit{large} otherwise. 
    
    To evaluate a model's performance improvement compared to a baseline, we define a metric called \textit{normalized AUC improvement}. This metric measures the extent of improvement achieved by the model relative to the maximum possible improvement. It is calculated using the following formula:
    \vspace{-0.1cm}
    \begin {align*}
            & \scriptstyle Normalized\ AUC\ improvement = \frac{AUC\ of\ Proposed Model - AUC\ of\ Baseline Model}{1 - AUC\ of\ Baseline Model} \times 100 \%  
    \end{align*}
    Here, 1 represents the highest possible AUC score. For example, to calculate the normalized AUC improvement of a model M over baseline B, where the AUC of M is 0.81 and the AUC of B is 0.75, we would compute (0.81-0.75) / (1 - 0.75) $\times$ 100 \% = 24\%. This result indicates that M's improvement is 24\% of the maximum possible AUC improvement. Since we build 100 models for each studied system (i.e., one for each bootstrap sample), we report the average normalized AUC improvement when comparing two models.

    We also rank the studied prediction models (KUM , TM , PROD, PROC, OWN) using Scott-Knott ESD (SK-ESD)~\citep{ghotra_feature_importance_ICSME, mittas_ranking_feature_TSE, kla_model_validation}. The SK-ESD is a method of multiple comparison that leverages a hierarchical clustering to partition the set of treatment values (e.g., medians or means) into statistically distinct groups with non-negligible difference~\citep{mittas_ranking_feature_TSE}. We use Tim Menzies' implementation of the SK-ESD technique, which employs non-parametric statistical tests and the Cliff's Delta measure of effect size~\citep{tim_sk}.
\end{itemize}

\begin{table}[!t]
    \caption{The effect size and median AUC value of KUM , PROD, PROC, OWN and TM defect prediction models.} \label{tab:auc_defect_class_kucls}
    \resizebox{\columnwidth}{!}{
    \begin{tabular}{@{}lccccccccc@{}}
    \toprule
    \multicolumn{1}{c}{}                                   & \multicolumn{5}{c}{\textbf{Median AUC}}                                                 & \multicolumn{4}{c}{\textbf{Statistical Test (Effect Size)}}                                                                                                                                                                                                                        \\ \cmidrule(l){2-10} 
    \multicolumn{1}{c}{\multirow{-2}{*}{\textbf{Release}}} & \textbf{KUM } & \textbf{PROD} & \textbf{PROC} & \textbf{OWN} & \textbf{TM} & \textbf{\begin{tabular}[c]{@{}c@{}}KUM\\vs PROD\end{tabular}} & \textbf{\begin{tabular}[c]{@{}c@{}}KUM\\vs PROC\end{tabular}} & \textbf{\begin{tabular}[c]{@{}c@{}}KUM \\vs OWN\end{tabular}} & \textbf{\begin{tabular}[c]{@{}c@{}}KUM\\vs TM \end{tabular}} \\ \midrule
    activemq-5.0.0                                         & 0.83           & 0.80              & 0.81              & 0.84             & 0.89        & \cellcolor[HTML]{67FD9A}L(0.77)*                                     & \cellcolor[HTML]{67FD9A}L(0.63)*                                     & S(-0.28)*                                                           & L(-1.00)*                                                      \\
activemq-5.1.0                                         & 0.83           & 0.80              & 0.62              & 0.73             & 0.82        & \cellcolor[HTML]{67FD9A}L(0.76)*                                     & \cellcolor[HTML]{67FD9A}L(1.00)*                                     & \cellcolor[HTML]{67FD9A}L(0.99)*                                    & \cellcolor[HTML]{67FD9A}S(0.21)*                               \\
activemq-5.2.0                                         & 0.83           & 0.80              & 0.69              & 0.82             & 0.85        & \cellcolor[HTML]{67FD9A}L(0.75)*                                     & \cellcolor[HTML]{67FD9A}L(1.00)*                                     & N(0.07)                                     & L(-0.70)*                                                      \\
activemq-5.3.0                                         & 0.82           & 0.80              & 0.66              & 0.81             & 0.84        & \cellcolor[HTML]{67FD9A}L(0.64)*                                     & \cellcolor[HTML]{67FD9A}L(1.00)*                                     & \cellcolor[HTML]{67FD9A}S(0.29)*                                    & L(-0.58)*                                                      \\
activemq-5.8.0                                         & 0.85           & 0.83              & 0.68              & 0.74             & 0.86        & \cellcolor[HTML]{67FD9A}L(0.61)*                                     & \cellcolor[HTML]{67FD9A}L(1.00)*                                     & \cellcolor[HTML]{67FD9A}L(1.00)*                                    & M(-0.41)*                                                      \\
derby-10.2.1.6                                         & 0.74           & 0.77              & 0.78              & 0.79             & 0.86        & L(-0.73)*                                                            & L(-0.95)*                                                            & L(-0.99)*                                                           & L(-1.00)*                                                      \\
derby-10.3.1.4                                         & 0.84           & 0.82              & 0.71              & 0.75             & 0.86        & \cellcolor[HTML]{67FD9A}S(0.18)*                                     & \cellcolor[HTML]{67FD9A}L(1.00)*                                     & \cellcolor[HTML]{67FD9A}L(1.00)*                                    & L(-0.72)*                                                      \\
derby-10.5.1.1                                         & 0.83           & 0.82              & 0.60              & 0.75             & 0.85        & \cellcolor[HTML]{67FD9A}M(0.44)*                                     & \cellcolor[HTML]{67FD9A}L(1.00)*                                     & \cellcolor[HTML]{67FD9A}L(1.00)*                                    & L(-0.87)*                                                      \\
groovy-1.5.7                                           & 0.81           & 0.77              & 0.53              & 0.71             & 0.78        & \cellcolor[HTML]{67FD9A}M(0.38)*                                     & \cellcolor[HTML]{67FD9A}L(0.93)*                                     & \cellcolor[HTML]{67FD9A}L(0.68)*                                    & \cellcolor[HTML]{67FD9A}S(0.24)*                               \\
groovy-1.6.BETA\_1                                     & 0.82           & 0.80              & 0.72              & 0.75             & 0.87        & \cellcolor[HTML]{67FD9A}S(0.23)*                                     & \cellcolor[HTML]{67FD9A}L(0.79)*                                     & \cellcolor[HTML]{67FD9A}L(0.76)*                                    & L(-0.64)*                                                      \\
groovy-1.6.BETA\_2                                     & 0.85           & 0.89              & 0.75              & 0.77             & 0.91        & L(-0.5`)*                                                            & \cellcolor[HTML]{67FD9A}L(0.90)*                                     & \cellcolor[HTML]{67FD9A}L(0.83)*                                    & L(-0.64)*                                                      \\
hbase-0.94.0                                           & 0.71           & 0.78              & 0.64              & 0.76             & 0.83        & L(-0.9`)*                                                            & \cellcolor[HTML]{67FD9A}L(0.97)*                                     & L(-0.78)*                                                           & L(-1.00)*                                                      \\
hbase-0.95.0                                           & 0.71           & 0.69              & 0.67              & 0.70             & 0.82        & \cellcolor[HTML]{67FD9A}L(0.90)*                                     & \cellcolor[HTML]{67FD9A}L(0.82)*                                     & \cellcolor[HTML]{67FD9A}S(0.22)*                                    & L(-1.00)*                                                      \\
hbase-0.95.2                                           & 0.71           & 0.65              & 0.63              & 0.69             & 0.73        & \cellcolor[HTML]{67FD9A}L(0.69)*                                     & \cellcolor[HTML]{67FD9A}L(0.83)*                                     & M(-0.39)*                                                           & L(-0.95)*                                                      \\
hive-0.10.0                                            & 0.72           & 0.78              & 0.67              & 0.79             & 0.85        & L(-0.90)*                                                            & \cellcolor[HTML]{67FD9A}M(0.43)*                                     & L(-0.93)*                                                           & L(-1.00)*                                                      \\
hive-0.12.0                                            & 0.72           & 0.79              & 0.62              & 0.80             & 0.85        & L(-0.97)*                                                            & \cellcolor[HTML]{67FD9A}L(0.94)*                                     & L(-0.98)*                                                           & L(-1.00)*                                                      \\
hive-0.9.0                                             & 0.79           & 0.86              & 0.57              & 0.70             & 0.91        & L(-0.92)*                                                            & \cellcolor[HTML]{67FD9A}L(1.00)*                                     & \cellcolor[HTML]{67FD9A}L(0.98)*                                    & L(-1.00)*                                                      \\
jruby-1.1                                              & 0.86           & 0.82              & 0.66              & 0.83             & 0.84        & \cellcolor[HTML]{67FD9A}L(0.66)*                                     & \cellcolor[HTML]{67FD9A}L(1.00)*                                     & \cellcolor[HTML]{67FD9A}M(0.42)*                                    & \cellcolor[HTML]{67FD9A}M(0.49)*                               \\
jruby-1.4.0                                            & 0.83           & 0.82              & 0.67              & 0.73             & 0.85        & \cellcolor[HTML]{67FD9A}M(0.45)*                                     & \cellcolor[HTML]{67FD9A}L(1.00)*                                     & \cellcolor[HTML]{67FD9A}L(0.98)*                                    & M(-0.44)*                                                      \\
jruby-1.5.0                                            & 0.84           & 0.82              & 0.75              & 0.82             & 0.86        & \cellcolor[HTML]{67FD9A}M(0.45)*                                     & \cellcolor[HTML]{67FD9A}L(0.95)*                                     & \cellcolor[HTML]{67FD9A}M(0.39)*                                    & S(-0.28)*                                                      \\
jruby-1.7.0.preview1                                   & 0.84           & 0.81              & 0.51              & 0.76             & 0.85        & \cellcolor[HTML]{67FD9A}L(0.74)*                                     & \cellcolor[HTML]{67FD9A}L(1.00)*                                     & \cellcolor[HTML]{67FD9A}L(0.90)*                                    & M(-0.39)*                                                      \\
lucene-2.3.0                                           & 0.82           & 0.84              & 0.88              & 0.85             & 0.92        & M(-0.42)*                                                            & L(-0.81)*                                                            & L(-0.74)*                                                           & L(-1.00)*                                                      \\
lucene-2.9.0                                           & 0.78           & 0.78              & 0.68              & 0.69             & 0.82        & \cellcolor[HTML]{67FD9A}M(0.41)*                                     & \cellcolor[HTML]{67FD9A}L(1.00)*                                     & \cellcolor[HTML]{67FD9A}L(1.00)*                                    & L(-0.77)*                                                      \\
lucene-3.0.0                                           & 0.91           & 0.89              & 0.71              & 0.68             & 0.90        & \cellcolor[HTML]{67FD9A}L(0.81)*                                     & \cellcolor[HTML]{67FD9A}L(1.00)*                                     & \cellcolor[HTML]{67FD9A}L(1.00)*                                    & \cellcolor[HTML]{67FD9A}M(0.46)*                               \\
lucene-3.1.0                                           & 0.73           & 0.72              & 0.64              & 0.61             & 0.75        & \cellcolor[HTML]{67FD9A}S(0.27)*                                     & \cellcolor[HTML]{67FD9A}L(0.87)*                                     & \cellcolor[HTML]{67FD9A}L(0.96)*                                    & L(-0.56)*                                                      \\
wicket-1.3.0-beta2                                     & 0.82           & 0.80              & 0.74              & 0.70             & 0.83        & \cellcolor[HTML]{67FD9A}M(0.44)*                                     & \cellcolor[HTML]{67FD9A}L(0.94)*                                     & \cellcolor[HTML]{67FD9A}L(1.00)*                                    & M(-0.43)*                                                      \\
wicket-1.3.0-incubating-beta-1                         & 0.86           & 0.84              & 0.85              & 0.79             & 0.87        & \cellcolor[HTML]{67FD9A}M(0.13)*                                     & \cellcolor[HTML]{67FD9A}M(0.38)*                                     & \cellcolor[HTML]{67FD9A}L(0.62)*                                    & L(-0.66)*                                                      \\
wicket-1.5.3                                           & 0.81           & 0.77              & 0.53              & 0.70             & 0.79        & \cellcolor[HTML]{67FD9A}L(0.73)*                                     & \cellcolor[HTML]{67FD9A}L(1.00)*                                     & \cellcolor[HTML]{67FD9A}L(0.97)*                                    & \cellcolor[HTML]{67FD9A}M(0.39)*                               \\ \midrule

    Median                                                 & 0.82           & 0.80              & 0.67              & 0.75             & 0.85        & \multicolumn{1}{l}{}                                                 & \multicolumn{1}{l}{}                                                 & \multicolumn{1}{l}{}                                                & \multicolumn{1}{l}{}                                           \\ \bottomrule
    \multicolumn{10}{l}{\pbox{30cm} {1. The green cells indicate cases where KUM  outperforms a baseline model with a non-negligible effect size.}} \\
    \multicolumn{10}{l}{\pbox{30cm} {2. Cliff's Delta effect size: L -- Large, M -- Medium, S -- Small, and N -- Negligible}} \\
    \multicolumn{10}{l}{\pbox{30cm} {3. The statistical tests with a p-value less than 0.05 are marked with an asterisk (*) in the effect size results.}} 
    \end{tabular}
    }
    \end{table}

\smallskip \noindent \textbf{Findings.}\observation{The minimum AUC of KUM is above 0.70 (i.e., AUC > 0.70), indicating that KUs are a good candidate to predict post-release defects.} Table~\ref{tab:auc_defect_class_kucls} shows the effect size and median AUC value of the prediction models for each studied release. The KUM column refers to the AUC of the models built with the KU feature set. We observe that the AUCs of KUM are always above 0.70 (minimum is 0.71). In particular, the median AUC of KUM across all the studied releases is 0.82. A defect prediction model is typically considered suitable to be used in practice when its AUC is higher than or equal to 0.70 \citep{jiarpakdee2020featureselection}. This result highlights that our KU features are good candidates to predict post-release defects.

\begin{figure}[!h]
	\centering
	\includegraphics[width=0.9\linewidth]{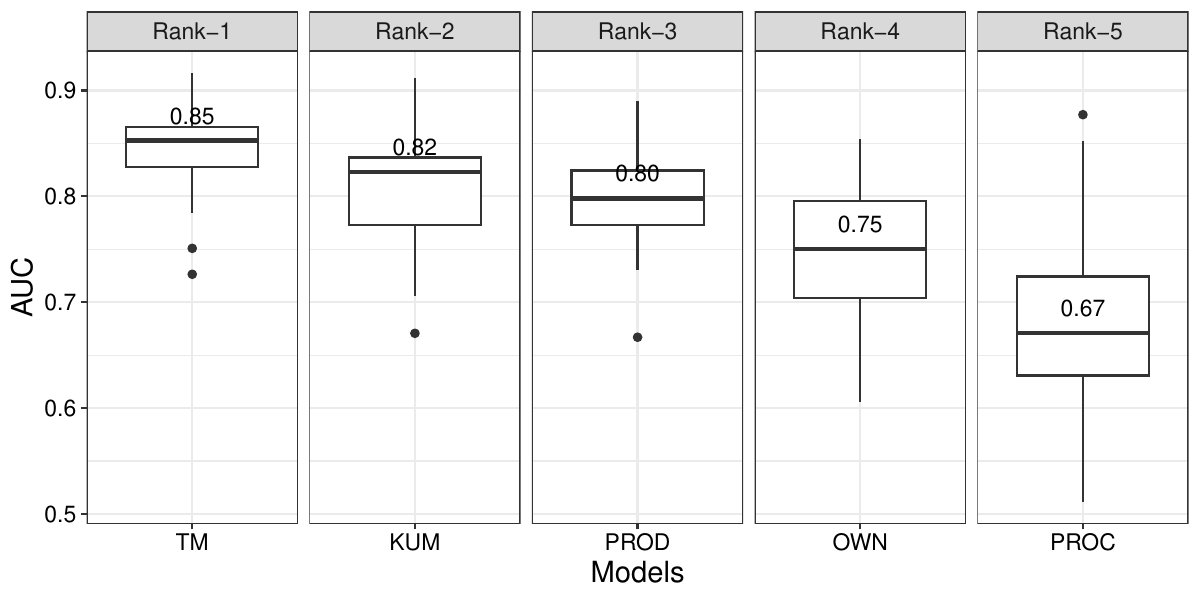}
	\caption{The distribution of AUC of KUM and other studied models. The models are grouped based on their performance rankings determined by the Scott-Knott ESD (SK-ESD) method, where a lower SK-ESD rank indicates a better-performing model.} 
	\label{fig:kucls_studied_mdoel_sk_rank}
\end{figure}

\smallskip \observation{KUM outperforms all baseline models except for TM.} Figure~\ref{fig:kucls_studied_mdoel_sk_rank} presents the SK-ESD ranks of the studied models across the studied releases. KUM ranks second, outperforming the baseline models that are built with one group of the traditional metrics (PROD, OWN and PROC). The median AUC values for these baseline models are 0.80 (PROD), 0.75 (OWN), and 0.67 (PROC), whereas, the median AUC for KUM is 0.82. Indeed, analysis of Table~\ref{tab:auc_defect_class_kucls} indicates that KUM consistently outperforms these three baseline models in the majority of the studied releases. For example, KUM outperforms PROD in 21 out of the 28 studied releases, typically with either a medium or large effect size (note green rows in Table~\ref{tab:auc_defect_class_kucls} of the column named ``KUM vs PROD''). KU metrics can thus be interpreted as a set of \textit{product} metrics with higher predictive power than traditional product metrics. Additionally, we calculate the normalized AUC improvement of KUM over PROD, PROC, and OWN for each release. The normalized AUC improvement of KUM over PROD ranges from 3.6\% to 22.2\% (median 11.4\%), improvement over PROC ranges from 6.7\% to 68.9\% (median 42.4\%), and improvement over OWN ranges from 3.3\% to 71.9\% (median 33.3\%). In summary, KUM not only outperforms a model built on the same raw data (PROD), but also those built on process (PROC) and ownership (OWN) data.

Finally, while TM does outperform KUM in 23 out of the 28 releases (Table~\ref{tab:auc_defect_class_kucls}), KUM achieves a median AUC of 0.82 in those 23 releases. This result indicates that KUM would still be a robust classifier for those releases in practice.

\begin{footnotesize}
    \begin{mybox}{Summary}
    	\textbf{RQ1: \RQOne}
        \tcblower
    	KUs are good candidates for predicting defect-prone code. In particular:
    	\begin{itemize}[itemsep = 0pt, label=\textbullet, wide = 0pt]
            \item KUM can classify post-release defects with a minimum AUC of 0.71 and a median AUC of 0.82. 
            \item KUM outperforms the baseline models that are built with each individual group of traditional metrics (PROC, PROD and OWN). The normalized AUC improvement of KUM over PROD ranges from 3.6\% to 22.2\% (median 11.4\%), improvement over PROC ranges from 6.7\% to 68.9\% (median 42.4\%), and improvement over OWN ranges from 3.3\% to 71.9\% (median 33.3\%).
            \item While KUM does not outperform TM (model built with all traditional metrics), KUM maintains strong AUC values.
    	\end{itemize}
    \end{mybox}
\end{footnotesize}
\subsection{RQ2: \RQTwo}
\label{sec:section_rq2}

\smallskip \noindent \textbf{Motivation.} Analyzing feature importances helps pinpoint the features that contribute most significantly to predicting post-release defects. This analysis not only enhances our understanding of the model's decision-making process but also highlights the practical relevance of specific KUs. 



\smallskip \noindent \textbf{Approach.} To determine the KU features that have the most significant impact on predictions, we conduct a model interpretation analysis utilizing SHAP (SHapley Additive exPlanation). SHAP is a widely adopted method for model interpretation that leverages game theory to provide an optimal framework for quantifying each feature's contribution to the model's predictions~\citep{molnar2020interpretable}. SHAP values can be positive or negative, signifying the direction of the feature's influence on the prediction. A positive SHAP value indicates that the feature increases the likelihood of the predicted outcome (e.g., classifying a file as defective in a defect prediction model). In contrast, a negative SHAP value means that the feature reduces the likelihood of the predicted outcome or increases the likelihood of the opposite class (i.e., clean class). The strength of a feature is determined by the magnitude of SHAP values. Specifically, the absolute SHAP value of a feature indicates its influence on the prediction, with higher values signifying greater importance. We use the \texttt{shapper} python package to perform the SHAP analysis \citep{shapper_package}. 

We compute feature importances for KUM as follows. First, we apply SHAP to KUM and calculate the absolute SHAP value for each feature across every file (i.e., record) in the dataset of a release. Next, for each feature, we sum the SHAP values of all files (i.e., all records) in the dataset. We follow this approach for all the 100 bootstraps of every release. As a result we obtain a distribution of the sum of SHAP values for every feature across all releases. Then, we apply the SK-ESD technique~\citep{mittas_ranking_feature_TSE} on the distribution of the sum of SHAP values generated from the 100 bootstraps of every release. As a result, we obtain a feature ranking for each release. Since there can be ties, two features with same importance are given the same rank. By taking all releases, we can produce a feature rank distribution for each feature. To summarize the results across all releases, we apply the SK-ESD technique again on the feature rank distributions, similarly to~\citet{ghotra_feature_importance_ICSME}. As a result, we obtain a final rank for each feature. 

\begin{figure}[!h]
	\centering
	\includegraphics[width=1.0\linewidth]{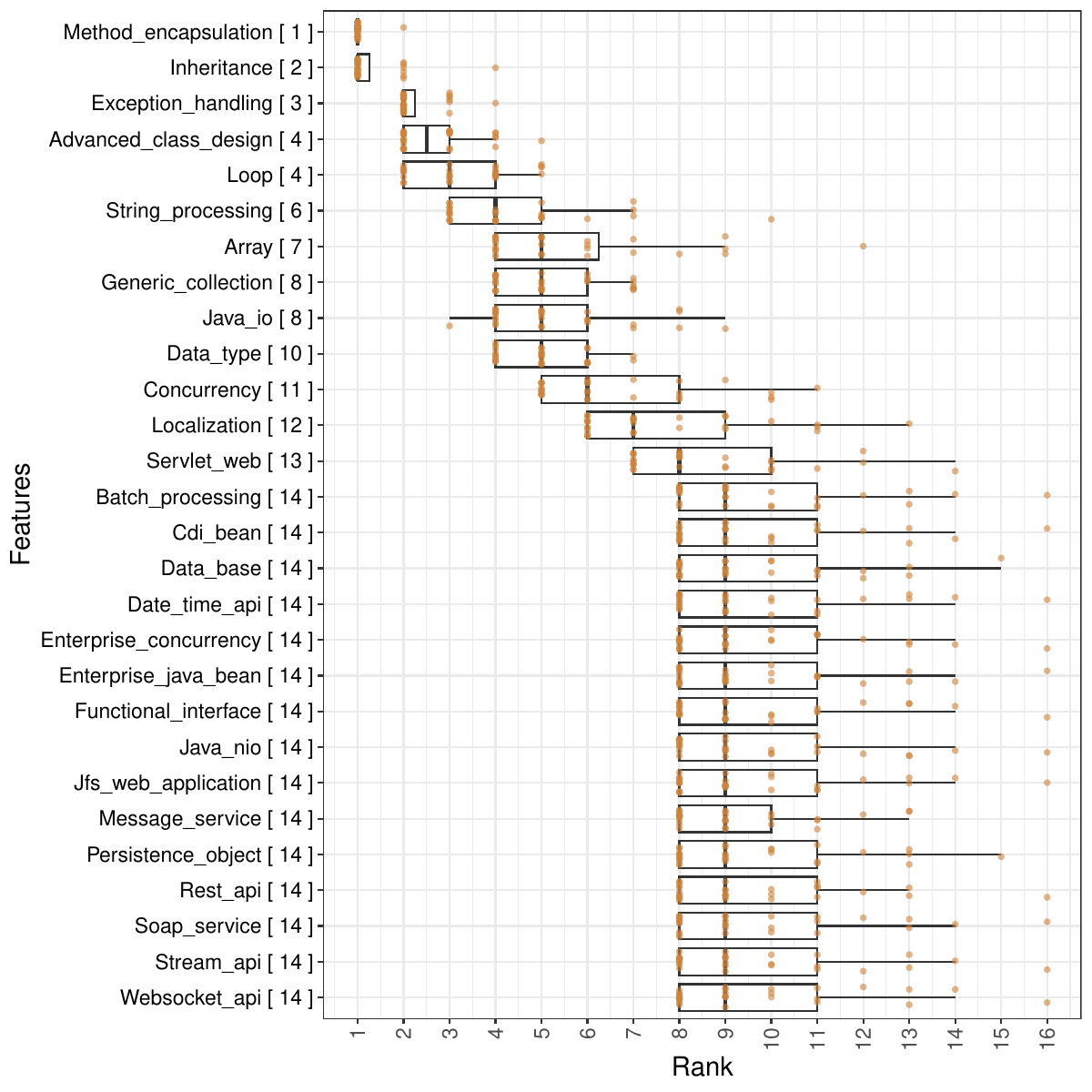}
	\caption{The rank distribution of features of KUM for defect classification. The number inside the square brackets indicates the final rank of the feature after applying Scott-Knott ESD for the second time.}
	\label{fig:kucsl_feature_importance}
\end{figure}

\smallskip \noindent \textbf{Findings.}\observation{Method \& Encapsulation is the most important KU feature across all studied releases of KUM for predicting post-release defects.}  Figure \ref{fig:kucsl_feature_importance} shows the rank distribution of each feature (final rank is shown in brackets). We observe that Method \& Encapsulation is the top-ranked feature in KUM. Indeed, the boxplot reveals that Method \& Encapsulation is the most important feature in most of our studied releases as the median value is one and the interquartile range is narrow (most ranks are one). Inheritance is ranked second and Exception Handling is the third-ranked feature. We then have two features ranked forth, namely: Advanced Class Design and Loop. Some advanced KUs derived from Java SE II certification exams also rank among the top ten most important features. For instance, Generics and Collections and Java IO are both ranked eighth. KUs related to building enterprise-level applications in Java, such as Enterprise JavaBeans and Java Message Service (JMS), which are derived from the Java EE certification exam, are ranked lower in importance (typically 14th). Upon closer inspection, we note that these advanced Java EE features are less frequently used (or even not used at all) in the studied releases. In contrast, fundamental KUs like Data Type and Loop, derived from the Java SE I exam, are more prevalent. 




\begin{footnotesize}
    \begin{mybox}{Summary}
    	\textbf{RQ2: \RQTwo}
        \tcblower
		Method \& Encapsulation, Inheritance, Exception Handling, Advanced Class Design, and Loop are the top five most important KU features for predicting post-release defects across all studied releases.
    \end{mybox}
\end{footnotesize}
\subsection{RQ3: \RQFour}
\label{sec:section_rq4}

\smallskip \noindent \textbf{Motivation.} This research question investigates whether the predictive performance of KUM for post-release defects can be improved by integrating them with traditional metrics. The underlying motivation is based on the hypothesis that combining diverse sets of features can provide a more comprehensive understanding of the factors influencing post-release defects. Such integration may enable the combined model to capture a wider range of predictive indicators, thereby enhancing its accuracy in identifying potential defects.

\smallskip \noindent \textbf{Approach.} \sloppy We build a combined model named KUM+TM for predicting post-release defects using all studied features of KUM and TM. Following Step 1 of RQ1 (Figure~\ref{fig:kucls_model_building}), we build KUM+TM models for all studied releases. We rank the studied prediction models (KUM, TM, PROD, PROC, OWN  and KUM+TM) using SK-ESD~\citep{ghotra_feature_importance_ICSME, mittas_ranking_feature_TSE, kla_model_validation}. We further compare the performance of KUM+TM model with the performance of KUM and TM models for each of the studied releases. We also analyze feature importances of KUM+TM model using the same SHAP analysis method explained in Section~\ref{sec:section_rq2}.

\begin{figure}[!t]
	\centering
	\includegraphics[width=0.9\linewidth]{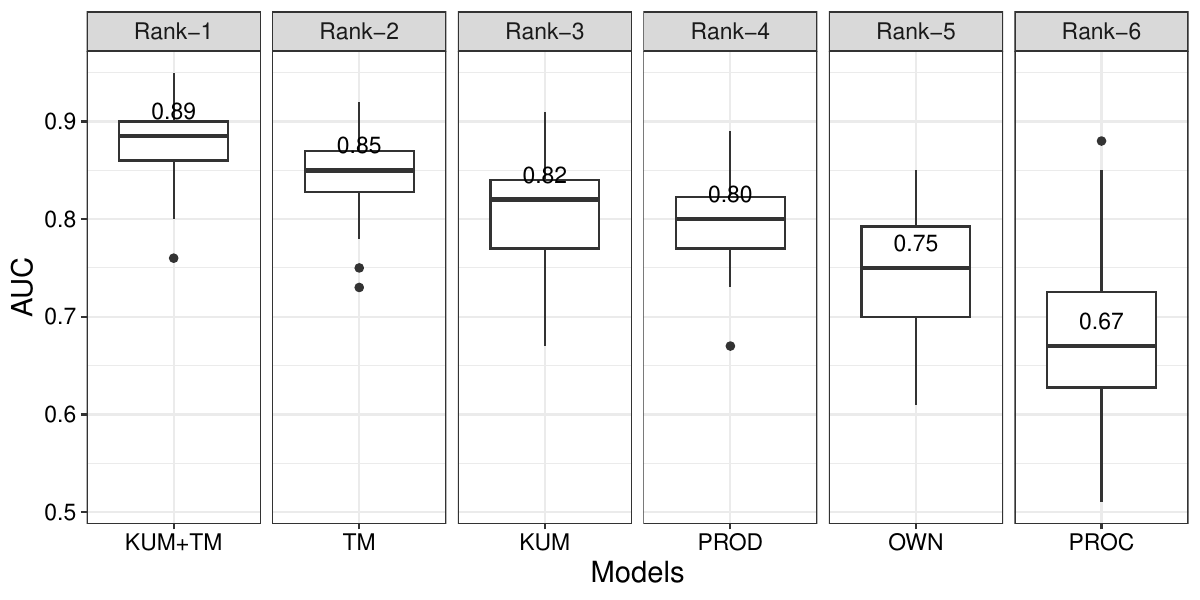}
	\caption{The distribution of AUC of the combined KUM+TM and other studied models. The models are grouped based on their performance rankings determined by the Scott-Knott ESD (SK-ESD) method, where a lower SK-ESD rank indicates a better-performing model.} 
	\label{fig:mdoel_sk_rank}
\end{figure}

\begin{table}[!h]
    \caption{The effect size and median AUC value of KUM, TM, and KUM+TM defect classification models.} \label{tab:auc_defect_class_combined}
    \resizebox{\columnwidth}{!}{
    \begin{tabular}{@{}lccccc@{}}
    \toprule
    \multicolumn{1}{c}{}                                   & \multicolumn{3}{c}{\textbf{Median AUC}}          & \multicolumn{2}{c}{\textbf{Statistical Test (Effect Size)}}         \\ \cmidrule(l){2-6} 
    \multicolumn{1}{c}{\multirow{-2}{*}{\textbf{Release}}} & \textbf{KUM+TM} & \textbf{KUM} & \textbf{TM} & \textbf{KUM+TM vs KUM}       & \textbf{KUM+TM vs TM}          \\ \midrule
    activemq-5.0.0                                         & 0.92              & 0.83           & 0.89        & \cellcolor[HTML]{67FD9A}L(1.00)* & \cellcolor[HTML]{67FD9A}L(0.94)* \\
    activemq-5.1.0                                         & 0.86              & 0.83           & 0.82        & \cellcolor[HTML]{67FD9A}L(0.68)* & \cellcolor[HTML]{67FD9A}L(0.81)* \\
    activemq-5.2.0                                         & 0.89              & 0.83           & 0.85        & \cellcolor[HTML]{67FD9A}L(0.98)* & \cellcolor[HTML]{67FD9A}L(0.85)* \\
    activemq-5.3.0                                         & 0.87              & 0.82           & 0.84        & \cellcolor[HTML]{67FD9A}L(0.97)* & \cellcolor[HTML]{67FD9A}L(0.81)* \\
    activemq-5.8.0                                         & 0.89              & 0.85           & 0.86        & \cellcolor[HTML]{67FD9A}L(0.97)* & \cellcolor[HTML]{67FD9A}L(0.93)* \\
    derby-10.2.1.6                                         & 0.89              & 0.74           & 0.86        & \cellcolor[HTML]{67FD9A}L(1.00)* & \cellcolor[HTML]{67FD9A}L(0.95)* \\
    derby-10.3.1.4                                         & 0.90              & 0.84           & 0.86        & \cellcolor[HTML]{67FD9A}L(1.00)* & \cellcolor[HTML]{67FD9A}L(0.99)* \\
    derby-10.5.1.1                                         & 0.89              & 0.83           & 0.85        & \cellcolor[HTML]{67FD9A}L(1.00)* & \cellcolor[HTML]{67FD9A}L(0.95)* \\
    groovy-1.5.7                                           & 0.83              & 0.81           & 0.78        & \cellcolor[HTML]{67FD9A}S(0.18)* & \cellcolor[HTML]{67FD9A}M(0.40)* \\
    groovy-1.6.BETA\_1                                     & 0.90              & 0.82           & 0.87        & \cellcolor[HTML]{67FD9A}L(0.81)* & \cellcolor[HTML]{67FD9A}M(0.42)* \\
    groovy-1.6.BETA\_2                                     & 0.95              & 0.85           & 0.91        & \cellcolor[HTML]{67FD9A}L(0.96)* & \cellcolor[HTML]{67FD9A}L(0.74)* \\
    hbase-0.94.0                                           & 0.86              & 0.71           & 0.83        & \cellcolor[HTML]{67FD9A}L(1.00)* & \cellcolor[HTML]{67FD9A}L(0.60)* \\
    hbase-0.95.0                                           & 0.84              & 0.71           & 0.82        & \cellcolor[HTML]{67FD9A}L(1.00)* & \cellcolor[HTML]{67FD9A}L(0.68)* \\
    hbase-0.95.2                                           & 0.76              & 0.67           & 0.73        & \cellcolor[HTML]{67FD9A}L(1.00)* & \cellcolor[HTML]{67FD9A}L(0.79)* \\
    hive-0.10.0                                            & 0.88              & 0.72           & 0.85        & \cellcolor[HTML]{67FD9A}L(1.00)* & \cellcolor[HTML]{67FD9A}L(0.87)* \\
    hive-0.12.0                                            & 0.88              & 0.72           & 0.85        & \cellcolor[HTML]{67FD9A}L(1.00)* & \cellcolor[HTML]{67FD9A}L(0.80)* \\
    hive-0.9.0                                             & 0.94              & 0.79           & 0.91        & \cellcolor[HTML]{67FD9A}L(1.00)* & \cellcolor[HTML]{67FD9A}L(0.96)* \\
    jruby-1.1                                              & 0.90              & 0.86           & 0.87        & \cellcolor[HTML]{67FD9A}L(0.58)* & \cellcolor[HTML]{67FD9A}L(0.63)* \\
    jruby-1.4.0                                            & 0.88              & 0.83           & 0.85        & \cellcolor[HTML]{67FD9A}L(0.84)* & \cellcolor[HTML]{67FD9A}L(0.67)* \\
    jruby-1.5.0                                            & 0.88              & 0.84           & 0.86        & \cellcolor[HTML]{67FD9A}L(0.68)* & \cellcolor[HTML]{67FD9A}L(0.54)* \\
    jruby-1.7.0.preview1                                   & 0.89              & 0.84           & 0.85        & \cellcolor[HTML]{67FD9A}L(0.93)* & \cellcolor[HTML]{67FD9A}L(0.83)* \\
    lucene-2.3.0                                           & 0.95              & 0.82           & 0.92        & \cellcolor[HTML]{67FD9A}L(1.00)* & \cellcolor[HTML]{67FD9A}L(0.91)* \\
    lucene-2.9.0                                           & 0.85              & 0.78           & 0.82        & \cellcolor[HTML]{67FD9A}L(0.99)* & \cellcolor[HTML]{67FD9A}L(0.88)* \\
    lucene-3.0.0                                           & 0.93              & 0.91           & 0.90        & \cellcolor[HTML]{67FD9A}L(0.67)* & \cellcolor[HTML]{67FD9A}L(0.91)* \\
    lucene-3.1.0                                           & 0.80              & 0.73           & 0.75        & \cellcolor[HTML]{67FD9A}L(0.96)* & \cellcolor[HTML]{67FD9A}L(0.74)* \\
    wicket-1.3.0-beta2                                     & 0.87              & 0.82           & 0.83        & \cellcolor[HTML]{67FD9A}L(0.93)* & \cellcolor[HTML]{67FD9A}L(0.81)* \\
    wicket-1.3.0-incubating-beta-1                         & 0.90              & 0.86           & 0.87        & \cellcolor[HTML]{67FD9A}L(0.95)* & \cellcolor[HTML]{67FD9A}L(0.80)* \\
    wicket-1.5.3                                           & 0.83              & 0.81           & 0.79        & \cellcolor[HTML]{67FD9A}M(0.41)* & \cellcolor[HTML]{67FD9A}L(0.72)* \\ \bottomrule
    Median                                                 & 0.89              & 0.82           & 0.85        & \multicolumn{1}{l}{}             & \multicolumn{1}{l}{}             \\ \bottomrule
    \multicolumn{6}{l}{\pbox{30cm} {1. The green color rows show cases where KUM+TM significantly outperform the studied models with a }} \\
    \multicolumn{6}{l}{\pbox{30cm} {non-negligible effect size.}} \\
    \multicolumn{6}{l}{\pbox{30cm} {2. Cliff’s Delta effect size: L -- Large, M -- Medium, S -- Small, and N -- Negligible}} \\
    \multicolumn{6}{l}{\pbox{30cm} {3. The statistical tests with a p-value less than 0.05 are marked with an asterisk (*) in the effect size results.}} 
    \end{tabular}
    }
    \end{table}

\smallskip \noindent \textbf{Findings.} \sloppy \observation{Combining KU features with traditional metrics results in a better performing model for predicting post-release defects.} Figure~\ref{fig:mdoel_sk_rank} depicts the performance of the studied models. We observe that the combined model KUM+TM is the top performing model. The median AUC of KUM+TM is 0.89, representing a normalized improvement of 36.4\% over TM and 63.6\% over KUM. We further examine the performance of KUM+TM for individual releases (Table~\ref{tab:auc_defect_class_combined}). We observe that KUM+TM models outperform TM and KUM in all 28 studied releases, with the majority showing a large effect size  (highlighted in green in the columns ``KUM+TM vs KUM'' and ``KUM+TM vs TM''). We also calculate the normalized AUC improvement for each of the studied releases. The normalized AUC improvement of KUM+TM over TM ranges from 11.1\% to 44.4\% (median 22.5\%), while the improvement over KUM ranges from 10.5\% to 72.2\% (median 31.5\%). We thus conclude that KUM+TM demonstrates superior performance for predicting post-release defects, making it more effective than either of the individual models.

\begin{figure}[!h]
	\centering
	\includegraphics[width=1.0\linewidth]{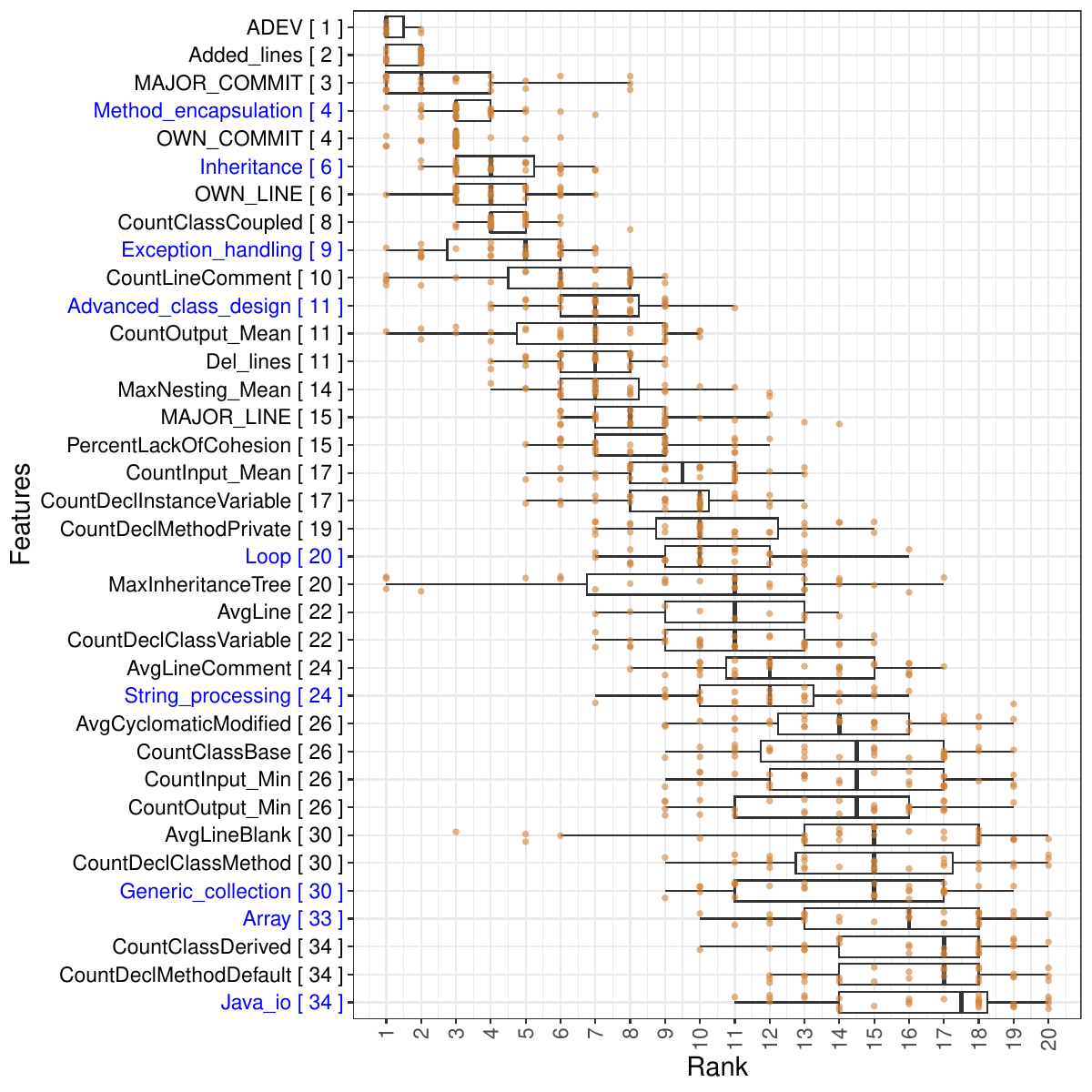}
	\caption{The rank distribution of features of KUM+TM. The number inside the square brackets indicates the final rank of the feature after applying Scott-Knott ESD for the second time. The KUs are highlighted with the \textcolor{blue}{blue} colored font and traditional metrics are highlighted with \textcolor{black}{black} colored font.}
	\label{fig:kucsl_cc_combined_feature_importance}
\end{figure}

\smallskip \sloppy\observation{ADEV is the top ranked feature for predicting post-release defects. Method \& Encapsulation is the highest ranked KU feature (fourth).} Figure~\ref{fig:kucsl_cc_combined_feature_importance} presents the rank distribution of each feature (final rank is shown in brackets) of KUM+TM across the studied releases. The top two most important features are process metrics: ADEV (i.e., the number of active developers) and Added\_Lines (the normalized count of lines added to the file). Major Commit, an ownership metric, is ranked third. We then have two features that are ranked as the forth most important feature. One of these features is Method \& Encapsulation, which is a KU feature. The other KU features in the top ten set are Inheritance (sixth) and Exception Handling (ninth). 

\begin{footnotesize}
    \begin{mybox}{Summary}
    	\textbf{RQ3: \RQFour}
        \tcblower
    	Yes. KUM+TM model outperforms both TM and KUM with a non-negligible effect size in every studied system. In particular:
    	\begin{itemize}[itemsep = 0pt, label=\textbullet, wide = 0pt]
            \item KUM+TM achieves an impressive median AUC of 0.89. 
            \item The normalized AUC improvement of KUM+TM over TM ranges from 11.1\% to 44.4\% (median 22.5\%) and the improvement over KUM ranges from 10.5\% to 72.2\% (median 31.5\%) across the studied releases.  
            \item In KUM+TM, the the number of active developers (ADEV) is the most important feature. Method \& Encapsulation is the top-ranked KU feature (fourth).
    	\end{itemize}
    \end{mybox}
\end{footnotesize}
\subsection{RQ4: \RQSix}
\label{sec:section_rq6}

\smallskip \noindent \textbf{Motivation.} Building predictive models with a broad set of features often yields high performance, but this approach can be resource-intensive, requiring significant efforts in data collection and processing. These challenges highlight the importance of developing cost-effective models that achieve strong predictive performance while relying on a minimal, carefully selected set of features. Such models not only reduce resource requirements but also enhance practicality and scalability in real-world.

\smallskip \noindent \textbf{Approach.} We focus on a subset of KU metrics and traditional metrics to build a cost-effective model. Specifically, we select the top five KU features (out of 28) from KUM and the top five metrics (out of 65) from TM. This results in the construction of our cost-effective model (COST\_EFF) using only 10 features  (a reduction of approximately 89\% in the number of features compared to KUM+TM). The selected features are as follows: Method \& Encapsulation KU, Exception KU, Advanced Class Design KU, Inheritance, Loop KU, ADEV, Added Lines, Major Commit, CountClassCoupled and CountLineComment.
    
To build COST\_EFF models, we follow the approach for model construction presented in Section~\ref{sec:section_rq1} (Figure~\ref{fig:kucls_model_building}). We then evaluate COST\_EFF and compare its performance to that of KUM, TM and KUM+TM. The comparison involves ranking the models according to the same SK-ESD method detailed in Section~\ref{sec:section_rq2}.

\begin{figure}[!t]
	\centering
	\includegraphics[width=1.0\linewidth]{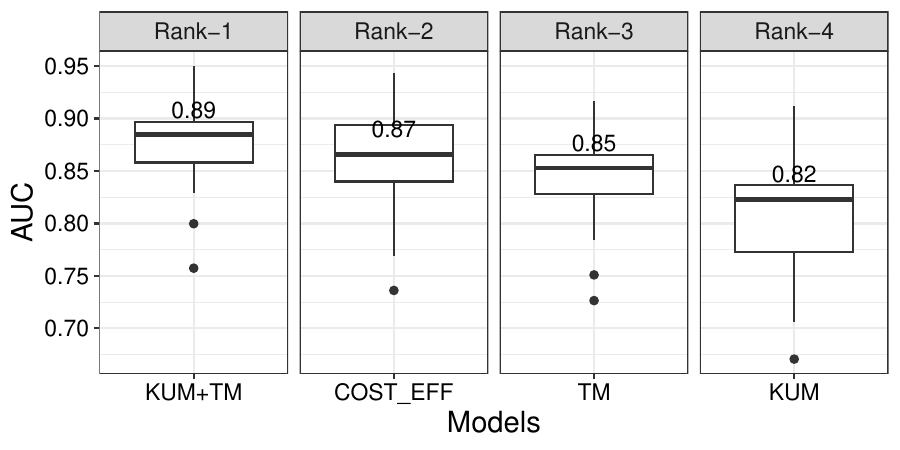}
	\caption{The distribution of AUC of COST\_EFF, KUM, TM, and KUM+TM across the studied releases. The models are grouped based on their performance rankings determined by the Scott-Knott ESD (SK-ESD) method, where a lower SK-ESD rank indicates a better-performing model.} 
	\label{fig:mdoel_sk_rank_cost}
\end{figure}
\begin{table}[!t]
    \caption{The effect size and median AUC value of COST\_EFF, TM, KUM and KUM+TM defect prediction models.} \label{tab:auc_defect_class_combined_cost}
    \resizebox{\columnwidth}{!}{
    \begin{tabular}{@{}lccccccc@{}}
    \toprule
    \multicolumn{1}{c}{}                                   & \multicolumn{4}{c}{\textbf{Median AUC}}                               & \multicolumn{3}{c}{\textbf{Statistical Test (Effect Size)}}                                                                                                                                                           \\ \cmidrule(l){2-8} 
    \multicolumn{1}{c}{\multirow{-2}{*}{\textbf{Release}}} & \textbf{COST\_EFF} & \textbf{TM} & \textbf{KUM} & \textbf{KUM+TM} & \textbf{\begin{tabular}[c]{@{}c@{}}COST\_EFF\\ vs TM\end{tabular}} & \textbf{\begin{tabular}[c]{@{}c@{}}COST\_EFF\\ vs KUM\end{tabular}} & \textbf{\begin{tabular}[c]{@{}c@{}}COST\_EFF\\ vs KUM+TM\end{tabular}} \\ \midrule
    activemq-5.0.0                                         & 0.91               & 0.89        & 0.83           & 0.92              & \cellcolor[HTML]{67FD9A}L(0.77)*                                   & \cellcolor[HTML]{67FD9A}L(1.00)*                                      & L(-0.58)*                                                                \\
    activemq-5.1.0                                         & 0.84               & 0.82        & 0.83           & 0.86              & \cellcolor[HTML]{67FD9A}L(0.53)*                                   & \cellcolor[HTML]{67FD9A}M(0.33)*                                      & M(-0.43)*                                                                \\
    activemq-5.2.0                                         & 0.88               & 0.85        & 0.83           & 0.89              & \cellcolor[HTML]{67FD9A}L(0.73)*                                   & \cellcolor[HTML]{67FD9A}L(0.97)*                                      & S(-0.25)*                                                                \\
    activemq-5.3.0                                         & 0.86               & 0.84        & 0.82           & 0.87              & \cellcolor[HTML]{67FD9A}L(0.52)*                                   & \cellcolor[HTML]{67FD9A}L(0.88)*                                      & L(-0.49)*                                                                \\
    activemq-5.8.0                                         & 0.87               & 0.86        & 0.85           & 0.89              & \cellcolor[HTML]{67FD9A}M(0.43)*                                   & \cellcolor[HTML]{67FD9A}L(0.72)*                                      & L(-0.81)*                                                                \\
    derby-10.2.1.6                                         & 0.87               & 0.86        & 0.74           & 0.89              & \cellcolor[HTML]{67FD9A}M(0.41)*                                   & \cellcolor[HTML]{67FD9A}L(1.00)*                                      & L(-0.84)*                                                                \\
    derby-10.3.1.4                                         & 0.83               & 0.86        & 0.84           & 0.90              & L(-0.84)*                                                          & S(-0.23)*                                                             & L(-1.00)*                                                                \\
    derby-10.5.1.1                                         & 0.85               & 0.85        & 0.83           & 0.89              & N(0.07)*                                                           & \cellcolor[HTML]{67FD9A}L(0.66)*                                      & L(-0.98)*                                                                \\
    groovy-1.5.7                                           & 0.79               & 0.78        & 0.81           & 0.83              & \cellcolor[HTML]{67FD9A}S(0.17)*                                   & S(-0.16)*                                                             & S(-0.31)*                                                                \\
    groovy-1.6.BETA\_1                                     & 0.90               & 0.87        & 0.82           & 0.90              & \cellcolor[HTML]{67FD9A}L(0.52)*                                   & \cellcolor[HTML]{67FD9A}L(0.87)*                                      & N(0.06)                                                                  \\
    groovy-1.6.BETA\_2                                     & 0.92               & 0.91        & 0.85           & 0.95              & \cellcolor[HTML]{67FD9A}S(0.22)*                                   & \cellcolor[HTML]{67FD9A}L(0.74)*                                      & L(-0.59)*                                                                \\
    hbase-0.94.0                                           & 0.86               & 0.83        & 0.71           & 0.86              & \cellcolor[HTML]{67FD9A}L(0.78)*                                   & \cellcolor[HTML]{67FD9A}L(1.00)*                                      & S(0.22)*                                                                 \\
    hbase-0.95.0                                           & 0.84               & 0.82        & 0.71           & 0.84              & \cellcolor[HTML]{67FD9A}L(0.75)*                                   & \cellcolor[HTML]{67FD9A}L(1.00)*                                      & N(0.08)                                                                  \\
    hbase-0.95.2                                           & 0.74               & 0.73        & 0.71           & 0.76              & \cellcolor[HTML]{67FD9A}S(0.33)*                                   & \cellcolor[HTML]{67FD9A}L(0.99)*                                      & L(-0.62)*                                                                \\
    hive-0.10.0                                            & 0.88               & 0.85        & 0.72           & 0.88              & \cellcolor[HTML]{67FD9A}L(0.89)*                                   & \cellcolor[HTML]{67FD9A}L(1.00)*                                      & N(0.05)                                                                  \\
    hive-0.12.0                                            & 0.88               & 0.85        & 0.72           & 0.88              & \cellcolor[HTML]{67FD9A}L(0.87)*                                   & \cellcolor[HTML]{67FD9A}L(1.00)*                                      & N(0.14)*                                                                 \\
    hive-0.9.0                                             & 0.92               & 0.91        & 0.79           & 0.94              & \cellcolor[HTML]{67FD9A}L(0.51)*                                   & \cellcolor[HTML]{67FD9A}L(1.00)*                                      & L(-0.73)*                                                                \\
    jruby-1.1                                              & 0.89               & 0.87        & 0.86           & 0.90              & \cellcolor[HTML]{67FD9A}L(0.55)*                                   & \cellcolor[HTML]{67FD9A}L(0.49)*                                      & N(-0.13)*                                                                \\
    jruby-1.4.0                                            & 0.85               & 0.85        & 0.83           & 0.88              & N(-0.11)*                                                          & \cellcolor[HTML]{67FD9A}M(0.38)*                                      & L(-0.74)*                                                                \\
    jruby-1.5.0                                            & 0.88               & 0.86        & 0.84           & 0.88              & \cellcolor[HTML]{67FD9A}L(0.58)*                                   & \cellcolor[HTML]{67FD9A}L(0.72)*                                      & N(0.04)                                                                  \\
    jruby-1.7.0.preview1                                   & 0.86               & 0.85        & 0.84           & 0.89              & \cellcolor[HTML]{67FD9A}S(0.29)*                                   & \cellcolor[HTML]{67FD9A}M(0.46)*                                      & L(-0.77)*                                                                \\
    lucene-2.3.0                                           & 0.94               & 0.92        & 0.82           & 0.95              & \cellcolor[HTML]{67FD9A}L(0.85)*                                   & \cellcolor[HTML]{67FD9A}L(1.00)*                                      & S(-0.32)*                                                                \\
    lucene-2.9.0                                           & 0.83               & 0.82        & 0.78           & 0.85              & \cellcolor[HTML]{67FD9A}M(0.38)*                                   & \cellcolor[HTML]{67FD9A}L(0.89)*                                      & L(-0.69)*                                                                \\
    lucene-3.0.0                                           & 0.90               & 0.90        & 0.91           & 0.93              & N(-0.02)                                                           & M(-0.47)*                                                             & L(-0.92)*                                                                \\
    lucene-3.1.0                                           & 0.77               & 0.75        & 0.72           & 0.80              & \cellcolor[HTML]{67FD9A}S(0.25)*                                   & \cellcolor[HTML]{67FD9A}L(0.72)*                                      & L(-0.61)*                                                                \\
    wicket-1.3.0-beta2                                     & 0.83               & 0.83        & 0.82           & 0.87              & N(-0.10)*                                                          & \cellcolor[HTML]{67FD9A}S(0.30)*                                      & L(-0.81)*                                                                \\
    wicket-1.3.0-incubating-beta-1                         & 0.90               & 0.87        & 0.84           & 0.90              & \cellcolor[HTML]{67FD9A}L(0.72)*                                   & \cellcolor[HTML]{67FD9A}L(0.93)*                                      & S(-0.16)*                                                                \\
    wicket-1.5.3                                           & 0.81               & 0.79        & 0.81           & 0.83              & \cellcolor[HTML]{67FD9A}M(0.40)*                                   & N(-0.04)                                                              & L(-0.48)*                                                                \\ \midrule
    Median                                                 & 0.87               & 0.85        & 0.82           & 0.89              & \multicolumn{1}{l}{}                                               & \multicolumn{1}{l}{}                                                  & \multicolumn{1}{l}{}                                                     \\ \bottomrule
    \multicolumn{8}{l}{\pbox{30cm} {1. The green color rows show cases where COST\_EFF significantly outperform the studied models with a non-negligible effect size.}} \\
    \multicolumn{8}{l}{\pbox{30cm} {2. Cliff’s Delta effect size: L -- Large, M -- Medium, S -- Small, and N -- Negligible.}} \\
    \multicolumn{8}{l}{\pbox{30cm} {3. The statistical tests with a p-value less than 0.05 are marked with an asterisk (*) in the effect size results.}} 
    \end{tabular}
    }
    \end{table}

\smallskip \noindent \textbf{Findings.} \sloppy \observation{COST\_EFF outperforms both TM and KUM, achieving a median AUC of 0.87}. Figure~\ref{fig:mdoel_sk_rank_cost} presents the AUC distribution of COST\_EFF, KUM, TM, and KUM+TM. COST\_EFF achieves a median AUC of 0.87, ranking second overall across the studied releases (i.e., just 0.02 absolute points behind the KUM+TM model). We further analyze the model's performance for each of the studied releases. Table~\ref{tab:auc_defect_class_combined_cost} presents the effect size of the classification models for each studied release. COST\_EFF outperforms the TM in most of the studied releases (23 out of 28), with a majority of these cases showing a large effect size. In another four releases, the two models perform the same (negligible effect size). Similarly, COST\_EFF outperforms KUM in most releases. The normalized AUC improvement of COST\_EFF over TM ranges from 3.7\% to 31.3\% (median 11.1\%), while the improvement over KUM ranges from 5.6\% to 66.7\% (median 26.8\%).

\begin{figure}[!h]
    \centering
    \includegraphics[width=0.75\linewidth]{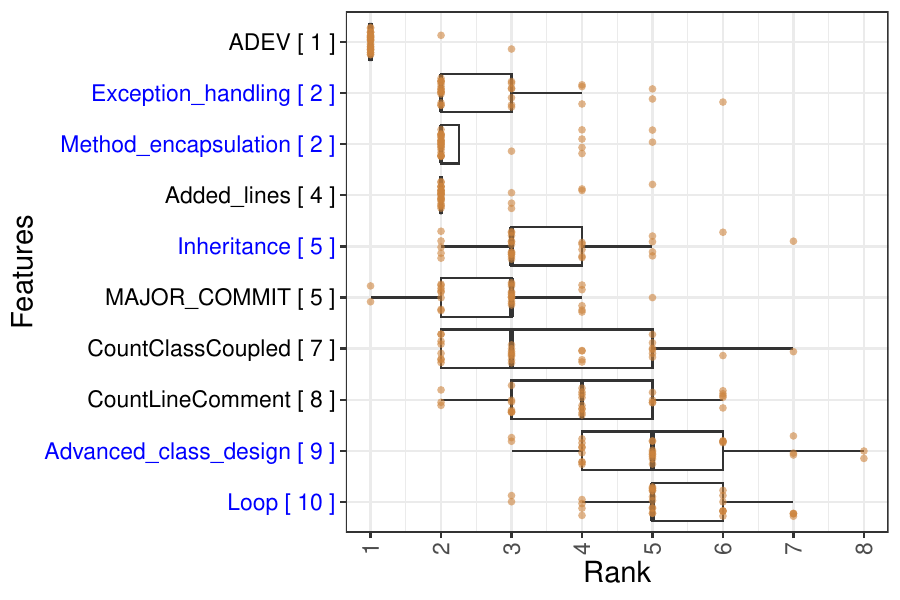}
    \caption{The rank distribution of features of COST\_EFF. The number inside the square brackets indicates the final rank of the feature after applying Scott-Knott ESD for the second time. The KUs are highlighted with the \textcolor{blue}{blue} colored font and traditional metrics are highlighted with \textcolor{black}{black} colored font.}
    \label{fig:kucsl_cc_cost_feature_importance}
    \end{figure}

\smallskip\noindent\sloppy\observation{Exception Handling and Method \& Encapsulation stand out as important features in COST\_EFF.} Figure~\ref{fig:kucsl_cc_cost_feature_importance} presents the rank distribution of each feature (final rank is shown in brackets) of COST\_EFF across the studied releases. Similar to the feature importance results of KUM+TM, the top most important feature of COST\_EFF is ADEV (the number of active developers). In particular, ADEV is consistently the top feature in almost all releases (note how the Q1, median, and Q3 all correspond to one). The KU features Exception Handling and Method \& Encapsulation are both ranked second. The narrow interquartile range of the Method \& Encapsulation feature, compared to the wider range of Exception Handling and other lower-ranked features, indicates that Method \& Encapsulation has a consistent ranking across the studied releases. Since three KU features are included in the top-5 ranked features, we conclude that KU features have a significant contribution in the cost-effective model.


\begin{footnotesize}
    \begin{mybox}{Summary}
    	\textbf{RQ4: \RQSix}
        \tcblower
        The cost-effective model, COST\_EFF,  outperforms TM (as well as KUM). In particular,
    	\begin{itemize}[itemsep = 0pt, label=\textbullet, wide = 0pt]
            \item The median AUC of COST\_EFF is 0.87.
            \item The normalized AUC improvement of COST\_EFF over TM ranges from 3.7\% to 31.3\% (median 11.1\%). 
            \item Similar to KUM+TM, ADEV is ranked as the most important feature. Method \& Encapsulation and Exception Handling are both ranked second, emphasizing the significant contribution of KU features to the performance of COST\_EFF.
    	\end{itemize}
    \end{mybox}
\end{footnotesize}
    \section{Discussion}
\label{sec:discussion}


In this section, we discuss the trade-offs between traditional metrics and KU metrics for predicting post-release defects and outline future directions for expanding the concept of KUs to include domain-specific and technical knowledge.

\subsection{On the trade-offs between traditional metrics and KU metrics for predicting post-release defects}


\begin{figure}[!t]
	\centering
	\includegraphics[width=0.8\linewidth]{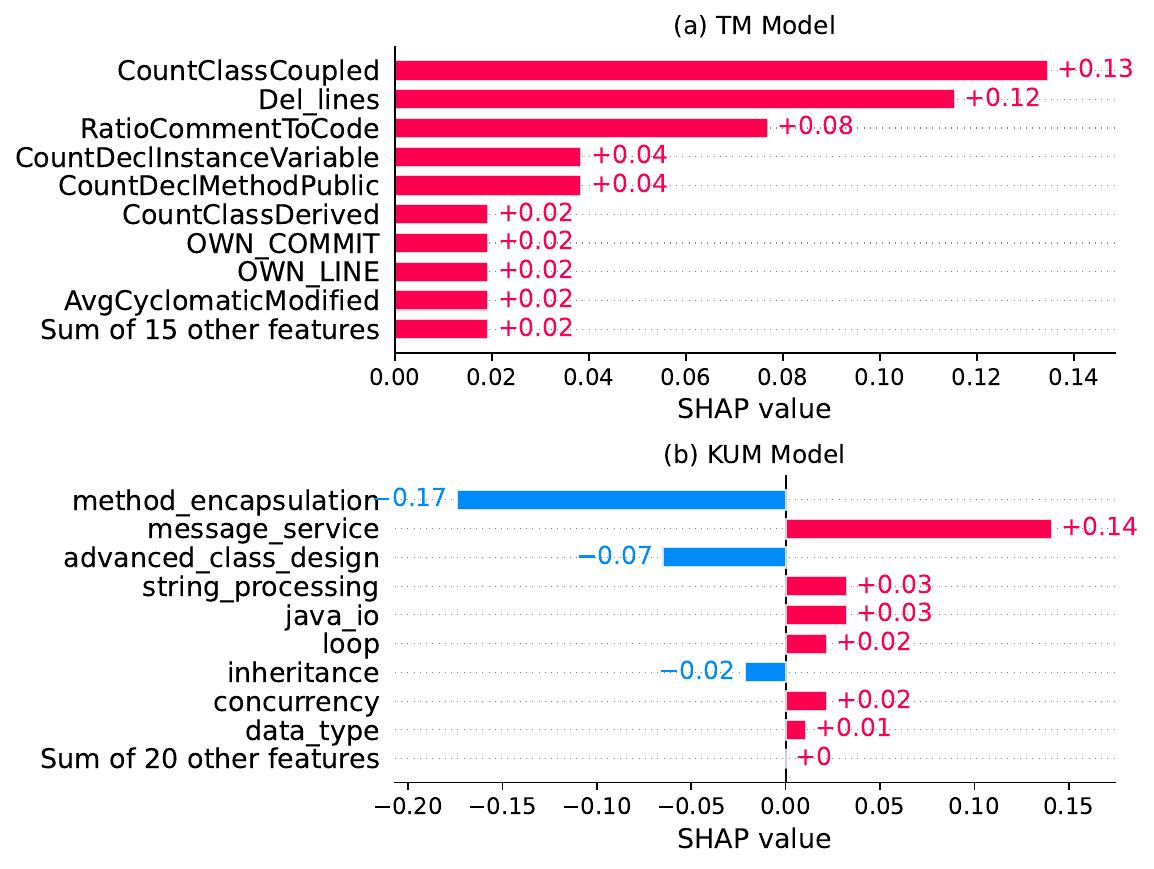}
	\caption{The SHAP values of the top ten features of the TM and KUM models for their contribution to classifying the ActiveMQSession.java file as defect-prone.}
	\label{fig:kucsl_cc_instance}
\end{figure}

Based on our findings, it is clear that KUs are \textbf{excellent} product metrics - even better than traditional product metrics (KUM outperforms PROD), which has been a long-standing standard in software engineering field. Furthermore, combining KUs with traditional metrics of TM leads to a top-performing model (KUM+TM) that achieves a median AUC of 0.89 and surpasses the already high median AUC of 0.85 achieved by TM. These findings demonstrate that KUs enhance the predictive power of traditional defect classification models. 

On the other hand, we acknowledge that the traditional metrics combo (process, product and ownership metrics) remains overall stronger than just KUs (i.e., TM outperforms KUM). Feature importance analysis of the combined model KUM+TM also reveals that traditional metrics still prevail in terms of feature importances.

This duality raises the question of which set of metrics one should use in practice. To answer this question, we would like to point out that actionability is as important as predictive power in several scenarios. Since KUs offer a new to interpret post-release defects (due to its unique capability of capturing system traits), we believe that KUs have the potential to offer unique actionability insights. As an illustrative example, consider the SHAP analysis of the ActiveMQSession.java file of the activemq-5.8.0 release for TM and KUM (Figure~\ref{fig:kucsl_cc_instance}). The SHAP for KUM indicates that Java Message Service API (message\_service) is a relevant and important KU that positively impacts the file classification as defective. By manually investigating the fixing patch\footnote{\url{https://shorturl.at/Kkc3L}}, we note that one of the changed methods (run(), line 860) does use Message Service APIs (an instantiation of an object of \texttt{MessageAck} class). In addition, the defect-prone class file instantiates and uses a set of methods and subclasses of several Message Service APIs (e.g., \texttt{Session}, \texttt{Message}, \texttt{QueueSender} and \texttt{QueueReceiver}). Therefore, it becomes clear that the defect-prone class (and the defect-prone method) do operate in a context where the concept of message service matters. In fact, the associated issue AMQ-4634\footnote{\url{https://issues.apache.org/jira/browse/AMQ-4634}} also mentions that the problem might be related to how ActiveMQ handles transactions in a Java Message Service (JMS) environment when working with distributed transactions. In contrast, the traditional metrics of TM (e.g., CountClassCoupled) do not clearly highlight the deeper insight about the context of the defect (see the SHAP for TM in Figure~\ref{fig:kucsl_cc_instance}(a)). Hence, we conclude that, in this case, KU features could provide further insights into the context in which defects happen. 

In summary, in the context where predictive power is the key factor, developers should opt for COST\_EFF, TM, or KUM+TM model, as those offer superior performance overall. However, if actionability and model interpretability are more important, developers may benefit from using KUM. Further research should be carried out to better understand the trade-offs between using traditional metrics and KUs (e.g., by investigating additional subject systems where Java EE KUs are more prevalent).

\subsection{The road ahead} 

\noindent \textbf{Application domains.} In this paper, we leverage KUs of the Java programming language to predict post-release defects. Our results clearly demonstrate that KUs are good features for defect prediction. Our prior work also showed that KUs can effectively enhance reviewer recommendation systems~\citep{ahasanuzzaman2024using}, further highlighting their versatility and potential in software engineering tasks. While our studies lay out a strong foundation for the use of KUs in software analysis, we see these contributions as the beginning of a broader research journey rather than an endpoint. A promising direction for future work would be to explore the application of KUs in supporting line-level defect prediction~\citep{fu2022linevul}. For instance, KUs could be mapped to line-level code features, such as the presence of specific constructs or the usage of critical APIs. This adaptation could provide a more granular understanding of defect-prone areas and further extend the applicability of KUs in fine-grained software system analysis.

\noindent \textbf{Alternative KU elicitation methods.} Our current approach to eliciting KUs and the associated feature engineering resulted in high-performing models for predicting post-release defects. However, the process of manually extracting KUs from certification exams is labor-intensive and does not easily scale to other programming languages or contexts. To address this limitation, researchers should explore more efficient and scalable methods for eliciting KUs. One promising avenue is leveraging Large Language Models (LLMs) to automate the elicitation and summarization of KUs~\citep{nam2024using}. LLMs could be trained or fine-tuned to extract KUs not only from certification exams but also from a broader range of resources, such as classic textbooks, programming tutorials, and domain-specific documentation. This approach could significantly expand the applicability of KUs across different languages and platforms. It is also worth noting that certification exams do exist for various programming languages and could serve as a foundation for eliciting KUs in those contexts. However, automating and generalizing the elicitation process by incorporating diverse sources would make the process more practical and adaptable for a wider range of programming languages and scenarios.

\noindent \textbf{Beyond KUs of programming languages.} Researchers should consider expanding KUs to include other forms of knowledge, such as domain-specific knowledge units. A promising approach could involve analyzing project descriptions from README files and developer discussions in pull requests or issue trackers, to extract relevant domain knowledge. By leveraging large language models (LLMs), this process can be automated and made scalable, enabling the identification and summarization of domain-specific concepts and their relationships to the software system. For example, Retrieval-Augmented Generation (RAG) architectures can be utilized, where LLMs are combined with retrieval mechanisms to dynamically pull relevant context from external knowledge sources, such as issues and pull request discussions. This allows the LLM to generate more accurate and contextually relevant summaries of domain knowledge (e.g., project-specific terminologies, workflows, or dependencies). We also encourage future works to integrate other form of technical knowledge related to libraries and APIs. This could include high-level libraries widely used in data science, such as pandas, as well as lower-level, such as \texttt{OpenSSL} for cryptographic operations, and \texttt{libcurl} for network communication. Expanding this scope could involve labeling issues with API domains~\citep{santos2023tag}, mapping analogous APIs between third-party libraries~\citep{liu2020generating}, and identifying API caveats or misuse patterns~\citep{li2018improving}. Incorporating these diverse forms of knowledge could provide deeper insights into code behavior and can significantly enhance the predictive power of models for identifying post-release defects.



    \section{Threats to Validity}
\label{sec:Limitations_And_Threats}




\noindent\textbf{Construct validity.} We use the Java certification exams as guidelines to identify KUs because certification exams cover important topics of the Java programming language. We only focus on the core functionalities of the programming language that are listed in the topics of these exams. The set of KUs that we conceive in this paper is thus limited and does not cover all the knowledge embedded in the constructs and APIs of the Java programming language. However, these core functionalities are covered in the Java certification exams (Java SE and Java EE) to assess the expertise of developers. We encourage future studies to refine our conceptualization of KUs and operationalize them in different ways.

Another construct validity threat stems from the fact that we did not extract KUs from the third-party libraries used by the studied systems. In this study, we only focus on those KUs that are implemented by the developers of the studied software systems. We carefully implemented the type binding resolutions using the Java JDT \citep{eclipse_jdt} to ensure that we are less likely to miss the KUs implemented within the studied projects.

We reused the 65 traditional metrics provided in the defect dataset, which include 54 product metrics, 5 process metrics, and 6 code ownership metrics. While there are many traditional metrics available, the selected traditional metrics for our study are widely recognized and have been extensively utilized in prior research to build and analyze defect prediction models~\citep{cross_project_kmeans,Varela17, jiarpakdee2020featureselection, laaber2021predicting, esteves2020understanding}.

\smallskip \noindent\textbf{Internal validity.} In our study, we used a Random Forest (RF) classifier with default hyper-parameter settings as the standard configuration for building the studied models. However, 
model performance can vary significantly across different classifiers and hyperparameter settings since each classifier has its unique strengths and weaknesses. To address this concern, we analyzed the impact of different classifiers and hyper-parameter tuning on model performance. We experimented with a range of classifiers, including traditional methods (e.g., Decision Trees, KNN) and advanced techniques (e.g., LightGBM, XGBoost). Our results showed that RF with default settings consistently achieved the best performance, minimizing bias associated with classifier selection and hyper-parameter tuning. Details of this analysis are provided in \ref{appendix:hyperparameter}.

The results reported in this paper are based on models that were trained without applying data rebalancing techniques, such as SMOTE \citep{chawla2002smote}. To investigate whether class rebalancing techniques could improve model performance, we applied SMOTE and re-ran our evaluations. However, we observed no improvement in performance. Moreover, such class rebalancing techniques are not recommended when building explainable models, as they can introduce issues like concept drift and hinder model interpretability \citep{Chak_ICSE_SEIP}. Therefore, we do not use SMOTE in our study.

\smallskip \noindent \textbf{External validity.} In this study, we only considered Java projects. Future studies should broaden the scope of our study and investigate how our findings apply to software projects written in other popular programming languages. While the findings demonstrate cost-effectiveness within the context of the studied dataset, it is unclear whether this cost-effectiveness can be generalized to other defect datasets. Differences in characteristics such as dataset size, structure, or domain could influence the applicability of the results to broader contexts. However, it is worth noting that the defect dataset used in this study consists of multiple releases of eight well-maintained and real-world software systems, ensuring a realistic evaluation.
    \section{Related Work}
\label{sec:Related_Work}

A number of studies have focused on analyzing different metrics for defect prediction models \citep{md_briand_design_measure_quality_jss,md_emam_faulty_class_prediction_jss,md_zhou_design_metrics_severity_faults_TSE,gyimothy_TSE_metrics_fault_prediction,nachi_metrics_component_failure,zhang_relationship_line_code_defects,md_Olague_fault_agine_TSE,md_kanmani_neural_network_fault_IST,moser_change_static_code_metrics_defect_prediction,md_aggrawal_oo_metrics,md_burrows_coupling_fault, kumar_effectiveness_software_metrics}. In this section, we highlight a few of these studies that are more closely related to our study.

The simplest, easiest to extract, and most commonly used product metric for bug prediction is the lines of code (LOC). \citet{zhang_relationship_line_code_defects} investigated the relationship between LOC and bug in Eclipse and several NASA projects. The author observed that typical classification techniques based on LOC were able to predict defective components reasonably well. \citet{koru2008investigation,koru2008theory} observed a power-law relationship between LOC and bug-proneness, with the latter increasing at a slower rate. They also observed that smaller modules were proportionally more bug-prone compared with larger ones. Although LOC correlates with the number of bugs~\citep{ostrand2005predicting,zhang_relationship_line_code_defects,english2009fault}, some studies find no strong evidence that LOC is a good indicator of bugs~\citep{fenton2000quantitative,andersson2007replicated}. Many researchers investigated other product metrics based on code, such as object-oriented metrics and complexity metrics for predicting bugs. These metrics are commonly referred to as code metrics. \citet{gyimothy_TSE_metrics_fault_prediction} used eight object-oriented metrics and traditional code-size metrics (e.g., LOC) to detect bug-proneness in the Mozilla source code. Their results indicated that all code metrics (except the one that calculates the number of direct descendants of a class) were significant for bug prediction. \citet{nachi_metrics_component_failure} built prediction models using 18 complexity metrics along three dimensions (module, function, and class). Their evaluation results showed that complexity metrics were able to predict post-release defects with good accuracy, but no single set of metrics was applicable to all projects.

Many prior studies build bug-prediction models with process metrics and investigate the performance of the bug prediction task between models that are built with product metrics and process metrics. \citet{nachi_historical_process_product_metrics} investigated the utility of using process and product metrics to estimate post-release bug-proneness in Windows XP-SP1 and Windows 2003 Server operating systems. They observed that process and product metrics could be used to identify bug-prone modules at statistically significant levels. \citet {moser_change_static_code_metrics_defect_prediction} proposed 18 process metrics based on the change history of the files (i.e., change metrics). They studied the performance of bug prediction models built with these 18 process metrics and 31 traditional static code metrics. They observed that predictors based on process metrics outperformed predictors based on static code metrics. In contrast with the file-level bug prediction approach, \citet{giger2012method} explored bug prediction models at the method level. They developed prediction models using a combination of 15 process metrics and 9 source code metrics to assess their performance on 21 Java projects. Their findings indicated that process metrics alone were effective predictors and significantly outperformed source code metrics. However, combining both types of metrics did not enhance the models' classification performance. \citet{rahman2013and} built defect prediction models using 54 code metrics and 14 process metrics and evaluated their performance on 85 releases of 12 major open-source projects. The results of their study revealed that code metrics were generally less effective than process metrics.
 
\citet{majumder2022revisiting} carried out an extensive study analyzing 722K commits from 700 GitHub projects to compare the effectiveness of bug prediction models that use process metrics against those that use product metrics. Their findings indicated that process metrics are superior in predicting defects. However, \citet{dalla2021within} observed that product metrics are better than process metrics for identifying bugs in Infrastructure-as-Code (IaC) scripts, which are used in DevOps to manage and provision infrastructure through machine-readable files.

In this paper, we empirically study knowledge units (KUs) of programming languages, a new perspective for analyzing the source code of software systems. Our goal is to understand if we can leverage KUs for predicting post-release defects. Our findings indicate that KUs are good candidates for identifying defect-prone code and offer insights into the occurrence of defects or the context in which those defects happen.

    \section{Conclusion}
\label{sec:Conclusion}

In this paper, we present an empirical study on how to leverage KUs of a programming language to predict post-release defects. We develop a defect prediction model called KUM, which uses KU features. We compare the performance of KUM with that of models built with traditional metrics (PROC, PROD, OWN, and TM). KUM significantly outperforms PROD, PROC and OWN. Combining KUs and traditional metrics leads to an even a higher performing model (KUM+TM) that significantly outperforms both TM and KUM. We also observe that a cost-effective model (COST\_EFF) that combines the top features from KUM and TM yields a performance that is similar to KUM+TM. 

Our encouraging results highlights the relevance of KUs in the context of defect prediction. Given our promising findings from this study, we encourage researchers to further explore and expand on the concept of KUs. This includes developing more advanced conceptualizations and elicitation of KUs, as well as evaluating whether the insights from our study can be generalized to other software systems and programming languages. More generally, we believe that KUs provide a brand new lens for analyzing software systems and there is still much to uncover about the potential of KUs. We thus also encourage future research to explore diverse application areas where KUs may provide valuable insights, including software quality and maintenance.
	\section*{Declarations}
\label{sec:declarations}

\subsection*{\textbf{Data Availability Statement (DAS)}}

\noindent A supplementary material package is provided online in the following link:
\url{http://bit.ly/3GoSmHL}. The contents will be made available on a public GitHub
repository once the paper is accepted.

\subsection*{\textbf{Funding and/or Conflicts of interests/Competing interests}} 

\noindent The authors declared that they have no conflict of interest.

	\begin{footnotesize}
		\bibliographystyle{spbasic}      
		\bibliography{bib/references.bib}   
	\end{footnotesize}	

	\clearpage
	
	\appendix
	
	\normalsize
	\vspace{-1ex}
	\clearpage
\noindent\textbf{APPENDICES}
\vspace{3ex}
\section{Java Certification Exams and Knowledge Units}
\label{appendix:cert-exams}
Table~\ref{tab:ku-from-java-exams} summarizes our KUs and their capabilities.

\begin{table}[!b]
    \centering
    \caption{Knowledge Units derived from the Java SE 8 Programmer I Exam, Java SE 8 Programmer II Exam, and Java EE Developer Exam.}
    \label{tab:ku-from-java-exams}
    \resizebox{\textwidth}{!}{
    \begin{tabular}{p{2.5cm} p{4.5cm} p{12cm}} 
    \toprule
    \multicolumn{1}{c}{\textbf{Certification Exam}} & 
    \multicolumn{1}{c}{\textbf{Knowledge Unit (KU)}} & 
    \multicolumn{1}{c}{\textbf{Key Capabilities}} \\ 
    \midrule

    \multirow{6}{*}{Java SE I} & \textbf{[K1]} Data Type & 
    \textbf{[C1]} Declare and initialize different types of variables (e.g., primitive types, parameterized types, and arrays), including casting between primitive types. \\
    \cmidrule(lr){2-3}
    
    & \textbf{[K2]} Operator \& Decision &  
    \textbf{[C1]} Use Java operators (e.g., assignment and postfix operators); use parentheses to override operator precedence. \\  
    & & \textbf{[C2]} Test equality between strings and other objects using \texttt{==} and \texttt{equals()}. \\  
    & & \textbf{[C3]} Create and use \texttt{if}, \texttt{if-else}, and ternary constructs. \\  
    & & \textbf{[C4]} Use a \texttt{switch} statement. \\ \cmidrule(lr){2-3}
    
    & \textbf{[K3]} Array &
    \textbf{[C1]} Declare, instantiate, initialize and use a one-dimensional array \\
    & & \textbf{[C2]} Declare, instantiate, initialize and use a multi-dimensional array 
    \\ \cmidrule(lr){2-3}
    
    & \textbf{[K4]} Loop &
    \textbf{[C1]} Create and use \texttt{while} loops \\
    & & \textbf{[C2]} Create and use \texttt{for} loops, including the \texttt{enhanced for} loop \\
    & & \textbf{[C3]} Create and use \texttt{do-while} loops \\
    & & \textbf{[C4]} Use \texttt{break} statement \\
    & & \textbf{[C5]} Use \texttt{continue} statement
    \\ \cmidrule(lr){2-3}
    
    & \textbf{[K5]} Method \& Encapsulation &
    \textbf{[C1]} Create methods with arguments and return values \\
    & & \textbf{[C2]} Apply the ``static'' keyword to methods, fields, and blocks \\
    & & \textbf{[C3]} Create an overloaded method and overloaded constructor \\
    & & \textbf{[C4]} Create a constructor chaining (use ``this()'' method to call one constructor from another constructor \\
    & & \textbf{[C5]} Use variable length arguments in the methods \\
    & & \textbf{[C6]} Use different access modifiers (e.g., private and protected) other than ``default'' \\
    & & \textbf{[C7]} Apply encapsulation: identify set and get method to initialize any private class variables \\
    & & \textbf{[C8]} Apply encapsulation: Immutable class generation-final class and initialize private variables through the constructor
    \\ \cmidrule(lr){2-3}
    
    & \textbf{[K6]} Inheritance &
    \textbf{[C1]} Use basic polymorphism (e.g., a superclass refers to a subclass) \\
    & &\textbf{[C2] }Use polymorphic parameter (e.g., pass instances of a subclass or interface to a method) \\
    & & \textbf{[C3]} Create overridden methods \\
    & & \textbf{[C4]} Create ``abstract'' classes and ``abstract'' methods \\
    & & \textbf{[C5]} Create ``interface'' and implement the interface \\
    & & \textbf{[C6]} Use ``super()'' and the ``super'' keyword to access the members(e.g., fields and methods) of a parent class \\
    & & \textbf{[C7]} Use casting in referring a subclass object to a superclass object
    \\ \midrule
    \end{tabular}
    }
\end{table}

\begin{table}[!b]
    \centering
    \resizebox{\columnwidth}{!}{
    \begin{tabular}{p{1.5cm} p{2.5cm} p{12cm}} 
    \toprule
    \multicolumn{1}{c}{\textbf{Certification Exam}} & 
    \multicolumn{1}{c}{\textbf{Knowledge Unit (KU)}} & 
    \multicolumn{1}{c}{\textbf{Key Capabilities}} \\ 
    \midrule

    \multirow{12}{*}{Java SE II} & \textbf{[K7]} Advanced Class Design &
    \textbf{[C1]} Create inner classes, including static inner classes, local classes, nested classes, and anonymous inner classes \\
    & & \textbf{[C2]} Develop code that uses the final \\
    & & \textbf{[C3]} Use enumerated types including methods and constructors in an ``enum'' type \\
    & & \textbf{[C4]} Create singleton classes and immutable classes
    \\ \cmidrule(lr){2-3}
    
    & \textbf{[K8]} Generics \& Collection &
    \textbf{[C1]} Create and use a generic class \\
    & & \textbf{[C2]} Create and use \texttt{ArrayList}, \texttt{TreeSet}, \texttt{TreeMap}, and \texttt{ArrayDeque} 
    \textbf{[C3]} Use \texttt{java.util.Comparator} and \texttt{java.lang.Comparable} interfaces \\
    & & \textbf{[C4]} Iterate using forEach methods of List
    \\ \cmidrule(lr){2-3}

    & \textbf{[K9]} Functional Interface &
    \textbf{[C1]} Use the built-in interfaces included in the \texttt{java.util.function} packages such as \texttt{Predicate}, \texttt{Consumer}, \texttt{Function}, and \texttt{Supplier} \\
    & & \textbf{[C2]} Develop code that uses primitive versions of functional interfaces \\
    & & \textbf{[C3]} Develop code that uses binary versions of functional interfaces \\
    & & \textbf{[C4]} Develop code that uses the \texttt{UnaryOperator} interface
    \\ \cmidrule(lr){2-3}
    
    & \textbf{[K10]} Stream API &
    \textbf{[C1]} Develop code to extract data from an object using \texttt{peek()} and \texttt{map()} methods, including primitive versions of the \texttt{map()} method  \\
    & & \textbf{[C2]} Search for data by using search methods of the Stream classes, including \texttt{findFirst}, \texttt{findAny}, \texttt{anyMatch}, \texttt{allMatch}, \texttt{noneMatch} \\
    & & \textbf{[C3]} Develop code that uses the Optional class \\
    & & \textbf{[C4]} Develop code that uses Stream data methods and calculation methods \\
    & & \textbf{[C5]} Sort a collection using Stream API \\
    & & \textbf{[C6]} Save results to a collection using the collect method \\
    & & \textbf{[C7]} \texttt{UseflatMap()} methods in the Stream API
    \\ \cmidrule(lr){2-3}

    & \textbf{[K11]} Exception &
    \textbf{[C1]} Create a try-catch block \\
    & & \textbf{[C2]} Use catch, multi-catch, and finally clauses \\
    & & \textbf{[C3]} Use autoclose resources with a try-with-resources statement \\
    & & \textbf{[C4]} Create custom exceptions and autocloseable resources \\
    & & \textbf{[C5]} Create and invoke a method that throws an exception \\
    & & \textbf{[C6]} Use common exception classes and categories(such as \texttt{NullPointerException}, \texttt{ArithmeticException}, \texttt{ArrayIndexOutOfBoundsException}, \texttt{ClassCastException}) \\
    & & \textbf{[C5]} Use assertions
    \\ \cmidrule(lr){2-3}
    
    & \textbf{[K12]} Date Time API &
    \textbf{[C1]} Create and manage date-based and time-based events including a combination of date and time into a single object using \texttt{LocalDate}, \texttt{LocalTime}, \texttt{LocalDateTime}, \texttt{Instant}, \texttt{Period}, and \texttt{Duration} \\
    & & \textbf{[C2]} Formatting date and times values for using different timezones \\
    & & \textbf{[C3]} Create and manage date-based and time-based events using Instant, Period, Duration, and Temporal Unit \\
    & & \textbf{[C4]} Create and manipulate calendar data using classes from \texttt{java.time.LocalDateTime}, \texttt{java.time.LocalDate}, \texttt{java.time.LocalTime}, \texttt{java.time.format.DateTimeFormatter}, and \texttt{java.time.Period} 
    \\ \cmidrule(lr){2-3}
    
    & \textbf{[K13]} IO &
    \textbf{[C1]} Read and write data using the console \\
    & & \textbf{[C2]} Use \texttt{BufferedReader}, \texttt{BufferedWriter}, \texttt{File}, \texttt{FileReader}, \texttt{FileWriter}, \texttt{FileInputStream}, \texttt{FileOutputStream}, \texttt{ObjectOutputStream}, \texttt{ObjectInputStream}, and \texttt{PrintWriter} in the \texttt{java.io} package
    \\ \cmidrule(lr){2-3}
    
    & \textbf{[K14]} NIO &
    \textbf{[C1]} Use the Path interface to operate on file and directory paths \newline
    \textbf{[C2]} Use the Files class to check, read, delete, copy, move, and manage metadata a file or directory
    \\ \cmidrule(lr){2-3}
    
    & \textbf{[K15]} String Processing &
    \textbf{[C1]} Search, parse and build strings \\
    & & \textbf{[C2]} Manipulate data using the \texttt{StringBuilder} class and its methods \\
    & & \textbf{[C3]} Use regular expression using the \texttt{Pattern} and \texttt{Matcher} class \\
    & & \textbf{[C4]} Use string formatting
    \\ \cmidrule(lr){2-3}
    
    & \textbf{[K16]} Concurrency &
    \textbf{[C1]} Create worker threads using \texttt{Runnable}, \texttt{Callable} and use an \texttt{ExecutorService} to concurrently execute tasks \\
    & & \textbf{[C2]} Use \texttt{synchronized} keyword and \texttt{java.util.concurrent.atomic} package to control the order of thread execution \\
    & & \textbf{[C3]} Use \texttt{java.util.concurrent}   collections and classes including \texttt{CyclicBarrier} and \texttt{CopyOnWriteArrayList} \\
    & & \textbf{[C4]} Use parallel Fork/Join Framework
    \\ \cmidrule(lr){2-3}
    \multirow{2}{*}{} & \textbf{[K17]} Database &
    \textbf{[C1]} Describe the interfaces that make up the core of the JDBC API, including the \texttt{Driver}, \texttt{Connection}, \texttt{Statement}, and \texttt{ResultSet} interfaces \\
    & & \textbf{[C2]} Submit queries and read results from the database, including creating statements, returning result sets, iterating through the results, and properly closing result sets, statements, and connections
    \\ \cmidrule(lr){2-3}
    & \textbf{[K18]} Localization &
    \textbf{[C1]} Read and set the locale by using the Locale object \\
    & & \textbf{[C2]} Build a resource bundle for each locale and load a resource bundle in an application
    \\ \midrule
    \end{tabular}
    }
\end{table}

\begin{table}[!b]
    \centering
    \resizebox{\textwidth}{!}{
    \begin{tabular}{p{1.5cm} p{2.5cm} p{13cm}} 
    \toprule
    \multicolumn{1}{c}{\textbf{Certification Exam}} & 
    \multicolumn{1}{c}{\textbf{Knowledge Unit (KU)}} & 
    \multicolumn{1}{c}{\textbf{Key Capabilities}} \\ \midrule
    \multirow{10}{*}{Java EE} & \textbf{[K19]} Java Persistence &
    \textbf{[C1]} Create JPA Entity and Object-Relational Mappings (ORM) \\
    & & \textbf{[C2]} Use Entity Manager to perform database operations, transactions, and locking with JPA entities \\
    & & \textbf{[C3]} Create and execute JPQL statements \\ \cmidrule(lr){2-3}
    
    & \textbf{[K20]} Enterprise Java Bean &
    \textbf{[C1]} Create session EJB components containing synchronous and asynchronous business methods, manage the life cycle container callbacks, and use interceptors. \\
    & & \textbf{[C2]} Create EJB timers \\ \cmidrule(lr){2-3}

    & \textbf{[K21]} Java Message Service API &
    \textbf{[C1]} Implement Java EE message producers and consumers, including Message-Driven beans \\
    & & \textbf{[C2]} Use transactions with JMS API \\ \cmidrule(lr){2-3}
    
    & \textbf{[K22]} SOAP Web Service &
    \textbf{[C1]} Create SOAP Web Services and Clients using JAX-WS API  \\
    & & \textbf{[C2]} Create marshall and unmarshall Java Objects by using JAXB API \\ \cmidrule(lr){2-3}

    & \textbf{[K23]} Servlet &
    \textbf{[C1]} Create Java Servlet and use HTTP methods \\
    & & \textbf{[C2]} Handle HTTP headers, parameters, cookies \\
    & & \textbf{[C3]} Manage servlet life cycle with container callback methods and WebFilters \\ \cmidrule(lr){2-3}

    & \textbf{[K24]} Java REST API &
    \textbf{[C1]} Apply REST service conventions \\
    & & \textbf{[C2]} Create REST Services and clients using JAX-RS API \\ \cmidrule(lr){2-3}

    & \textbf{[K25]} Websocket &
    \textbf{[C1]} Create WebSocket Server and Client Endpoint Handlers \\
    & & \textbf{[C2]} Produce and consume, encode and decode WebSocket messages \\ \cmidrule(lr){2-3}

    & \textbf{[K26]} Java Server Faces &
    \textbf{[C1]} Use JSF syntax and use JSF Tag Libraries \\
    & & \textbf{[C2]} Handle localization and produce messages \\
    & & \textbf{[C3]} Use Expression Language (EL) and interact with CDI beans \\ \cmidrule(lr){2-3}
    
    & \textbf{[K27]} Contexts and Dependency Injection (CDI) &
    \textbf{[C1]} Create CDI Bean Qualifiers, Producers, Disposers, Interceptors, Events, and Stereotypes \\ \cmidrule(lr){2-3}

    & \textbf{[K28]} Batch Processing &
    \textbf{[C1]} Implement batch jobs using JSR API \\ \bottomrule

\end{tabular}
}
\end{table}

\section{Traditional code metrics} 
\label{appendix:code-metrics}

The list of process metrics employed in this study is presented in Table~\ref{tab:traditional_code_metrics}. Table~\ref{tab:process_metrics} presents the list of process metrics and Table~\ref{tab:owner_metrics} presents the list of ownership metrics that we study in this paper. These tables were adapted from the original paper of \citet{Yatish_post_release_defect}.

\begin{table}[!htbp]
\caption{The list of traditional code metrics that we select for this study.}
\label{tab:traditional_code_metrics}
\resizebox{\textwidth}{!}{
\begin{tabular}{@{}llll@{}}
\toprule
\multicolumn{1}{c}{\textbf{Granularity}} & \multicolumn{1}{c}{\textbf{Metric Name}} & \multicolumn{1}{c}{\textbf{Understand (API) name}} & \multicolumn{1}{c}{\textbf{Definition}}                                             \\ \midrule
File                           & Average Cyclomatic Complexity            & AvgCyclomatic                                      & Average cyclomatic complexity for all nested functions or methods.                  \\
                                    LeveL(37)     & Average Modified Cyclomatic Complexity   & AvgCyclomaticModified                              & Average modified cyclomatic complexity for all nested functions or methods.         \\
                                         & Average Strict Cyclomatic Complexity     & AvgCyclomaticStrict                                & Average strict cyclomatic complexity for all nested functions or methods.           \\
                                         & Average Essential Cyclomatic Complexity  & AvgEssential                                       & Average Essential complexity for all nested functions or methods.                   \\
                                         & Average Number of Lines                  & AvgLine                                            & Average number of lines for all nested functions or methods.                        \\
                                         & Average Number of Blank Lines            & AvgLineBlank                                       & Average number of blanks for all nested functions or methods.                       \\
                                         & Average Number of Lines of Code          & AvgLineCode                                        & Average number of lines containing source code for all nested functions or methods. \\
                                         & Average Number of Lines with Comments    & AvgLineComment                                     & Average number of lines containing comments for all nested functions or methods.    \\
                                         & Classes                                  & CountDeclClass                                     & Number of classes.                                                                  \\
                                         & Class Methods                            & CountDeclClassMethod                               & Number of class methods.                                                            \\
                                         & Class Variables                          & CountDeclClassVariable                             & Number of class variables.                                                          \\
                                         & Function                                 & CountDeclFunction                                  & Number of functions.                                                                \\
                                         & Instance MethodS(NIM)                   & CountDeclInstanceMethod                            & Number of instance methods.                                                         \\
                                         & Instance VariableS(NIV)                 & CountDeclInstanceVariable                          & Number of instance variables.                                                       \\
                                         & Local Methods                            & CountDeclMethod                                    & Number of local methods.                                                            \\
                                         & Local Default Visibility Methods         & CountDeclMethodDefault                             & Number of local default methods.                                                    \\
                                         & Private MethodS(NPM)                    & CountDeclMethodPrivate                             & Number of local private methods.                                                    \\
                                         & Protected Methods                        & CountDeclMethodProtected                           & Number of local protected methods.                                                  \\
                                         & Public Methods                           & CountDeclMethodPublic                              & Number of local public methods.                                                     \\
                                         & Physical LineS(NL)                      & CountLine                                          & Number of all lines.                                                                \\
                                         & Blank Lines of Code (BLOC)               & CountLineBlank                                     & Number of blank lines.                                                              \\
                                         & Source Lines of Code                     & CountLineCode                                      & Number of lines containing source code. {[}aka LOC{]}                               \\
                                         & Declarative Lines of Code                & CountLineCodeDecl                                  & Number of lines containing declarative source code.                                 \\
                                         & Executable Lines of Code                 & CountLineCodeExe                                   & Number of lines containing executable source code.                                  \\
                                         & Lines with Comments                      & CountLineComment                                   & Number of lines containing comments.                                                \\
                                         & Semicolons                               & CountSemicolon                                     & Number of semicolons.                                                               \\
                                         & Statements                               & CountStmt                                          & Number of statements.                                                               \\
                                         & Declarative Statements                   & CountStmtDecl                                      & Number of declarative statements.                                                   \\
                                         & Executable Statements                    & CountStmtExe                                       & Number of executable statements.                                                    \\
                                         & Max Cyclomatic Complexity                & MaxCyclomatic                                      & Maximum cyclomatic complexity of all nested functions or methods.                   \\
                                         & Max Modified Cyclomatic Complexity       & MaxCyclomaticModified                              & Maximum modified cyclomatic complexity of nested functions or methods.              \\
                                         & Max Strict Cyclomatic Complexity         & MaxCyclomaticStrict                                & Maximum strict cyclomatic complexity of nested functions or methods.                \\
                                         & Comment to Code Ratio                    & RatioCommentToCode                                 & Ratio of comment lines to code lines.                                               \\
                                         & Sum Cyclomatic Complexity (WMC)          & SumCyclomatic                                      & Sum of cyclomatic complexity of all nested functions or methods.                    \\
                                         & Sum Modified Cyclomatic Complexity       & SumCyclomaticModified                              & Sum of modified cyclomatic complexity of all nested functions or methods.           \\
                                         & Sum Strict Cyclomatic Complexity         & SumCyclomaticStrict                                & Sum of strict cyclomatic complexity of all nested functions or methods.             \\
                                         & Sum Essential Complexity                 & SumEssential                                       & Sum of essential complexity of all nested functions or methods                      \\ \midrule
Class                      & Coupling Between ObjectS(CBO)           & CountClassCoupled                                  & Number of other classes coupled to.                                                 \\
                                 LeveL(5)            & Number of ChildreN(NOC)                 & CountClassDerived                                  & Number of immediate subclasses.                                                     \\
                                         & Depth of Inheritance Tree (DIT)          & MaxInheritanceTree                                 & Maximum depth of class in inheritance tree.                                         \\
                                         & Lack of Cohesion in MethodS(LCOM)       & PercentLackOfCohesion                              & Lack of Cohesion in Methods                                                         \\
                                         & Response for clasS(RFC)                 & CountDeclMethodAll                                 & Number of methods, including inherited ones.                                        \\ \midrule
Method                        & InputS(FANINN)                          & CountInput \{Min, Median, Max\}                    & Number of calling subprograms plus global variables read.                           \\
                                LeveL(12)         & OutputS(FANOUT)                         & CountOutput \{Min, Median, Max\}                   & Number of called subprograms plus global variables set.                             \\
                                         & PathS(NPATH)                            & CountPath \{Min, Median, Max\}                     & Number of possible paths, not counting abnormal exits or gotos.                     \\
                                         & Nesting                                  & MaxNesting \{Min, Median, Max\}                    & Maximum nesting level of control constructs.                                        \\ \bottomrule
\end{tabular}
}
\end{table}

\clearpage
\begin{table}[]
    \caption{The list of process metrics that we select for this study.}
    \label{tab:process_metrics}
    \begin{tabular}{@{}ll@{}}
    \toprule
    Metric Name  & Description                                            \\ \midrule
    COMM         & The number of Git commits                              \\
    ADDED\_LINES & The normalized number of lines added to the file     \\
    DEL\_LINES   & The normalized number of lines deleted from the file \\
    ADEV         & The number of active developers                        \\
    DDEV         & The number of distinct developers                      \\ \bottomrule
\end{tabular}
\end{table}

\begin{table}[]
    \caption{The list of code ownership metrics that we select for this study.}
    \label{tab:owner_metrics}
    \resizebox{\textwidth}{!}{
        \begin{tabular}{p{4.0cm}p{14cm}}
    \toprule
    Metric Name  & Description                                            \\ \midrule
    MINOR\_COMMIT         & The number of unique developers who have contributed less than 5\% of the total code changes (i.e., Git commits) on the file                              \\
    MINOR\_LINE & The number of unique developers who have contributed less than 5\% of the total lines of code on the file     \\
    MAJOR\_COMMIT   & The number of unique developers who have contributed more than 5\% of the total code changes (i.e., Git commits) on the file \\
    MAJOR\_LINE         & The number of unique developers who have contributed more than 5\% of the total lines of code on the file                      \\
    OWN\_COMMIT         & The proportion of code changes (i.e., Git commits) made by the developer who has the highest contribution of code changes on the file\\
    OWN\_LINE         & The proportion of lines of code written by the developer who has the highest contribution of lines of code on the file developers                      \\ \bottomrule
\end{tabular}
    }
\end{table}

\section{On the influence of classifier choice and hyper-parameter tuning}\label{appendix:hyperparameter}


\begin{table}[!t]
    \caption{The studied classification algorithms with different parameter settings for hyper parameter optimization.}
    \label {tab:hyper_parameter_config}
    \resizebox{\columnwidth}{!}{
    \begin{tabular}{@{}lllll@{}}
    \toprule
    \textbf{\begin{tabular}[c]{@{}l@{}}Classifier\\ Name\end{tabular}} & \textbf{\begin{tabular}[c]{@{}l@{}}Classification\\ Algorithm\end{tabular}} & \textbf{\begin{tabular}[c]{@{}l@{}}Parameter\\ Name\end{tabular}} & \textbf{\begin{tabular}[c]{@{}l@{}}Parameter\\ Description\end{tabular}}                                                                                                                  & \textbf{\begin{tabular}[c]{@{}l@{}}Studied candidate\\ parameter values\end{tabular}} \\ \midrule
    KNN                                                                & \begin{tabular}[c]{@{}l@{}}K-Nearest \\ Neighbour\end{tabular}              & n\_neighbors                                                      & \begin{tabular}[c]{@{}l@{}}The number of neighbors  \\ required for  each sample\end{tabular}                                                                                             & \{1, 5, 9, 13, 17, 20\}                                                               \\ \midrule
    NB                                                                 & Naive Bayes                                                                 & var\_smoothing                                                    & \begin{tabular}[c]{@{}l@{}}The portion of the largest variance \\ of all features  that is added to \\ variances  for calculation stability.\end{tabular}                                 & \{1e-5, 1e-9, 1e-11, 1e-15\}                                                          \\ \midrule
    \multirow{3}{*}{DT}                                                & \multirow{3}{*}{Decision Tree}                                              & criterion                                                         & \begin{tabular}[c]{@{}l@{}}The function to measure the quality \\ of a split.\end{tabular}                                                                                                & \{`gini', `entropy', `log\_loss'\}                                                    \\ \cmidrule{3-5}
                                                                       &                                                                             & max\_depth                                                        & The maximum depth of the tree.                                                                                                                                                            & \{None, 5, 10\}                                                                       \\ \cmidrule{3-5}
                                                                       &                                                                             & ccp\_alpha                                                        & \begin{tabular}[c]{@{}l@{}}Complexity parameter used for \\ Minimal  Cost-Complexity Pruning.\end{tabular}                                                                                & \{0.0001, 0.001, 0.01, 0.1, 0.5\}                                                     \\ \midrule
    \multirow{2}{*}{RF}                                                & \multirow{2}{*}{Random Forest}                                              & n\_estimator                                                      & The number of trees in the forest.                                                                                                                                                        & \{10, 50, 100, 200\}                                                                  \\ \cmidrule{3-5}
                                                                       &                                                                             & max\_depth                                                        & \begin{tabular}[c]{@{}l@{}}The maximum depth of the tree. \\ If None,  then  nodes  are expanded \\ until all leaves  are pure.\end{tabular}                                              & \{None, 5, 10\}                                                                       \\ \midrule
    \multirow{3}{*}{XGB}                                               & \multirow{3}{*}{XGBoost}                                                    & n\_estimator                                                      & \begin{tabular}[c]{@{}l@{}}The number of boosting rounds or \\ trees  to build.\end{tabular}                                                                                              & \{10, 50, 100, 200\}                                                                  \\ \cmidrule{3-5}
                                                                       &                                                                             & max\_depth                                                        & The maximum depth of the tree.                                                                                                                                                            & \{None, 5, 10\}                                                                       \\ \cmidrule{3-5}
                                                                       &                                                                             & learning\_rate                                                    & \begin{tabular}[c]{@{}l@{}}This parameter controls how much \\ the  model is  adjusted  in response\\ to the  estimated error each  time \\ the model   weights are updated.\end{tabular} & \{0.1, 0.01, 0.001\}                                                                  \\ \midrule
    \multirow{3}{*}{LGBM}                                              & \multirow{3}{*}{LightGBM}                                                   & n\_estimator                                                      & The number of boosted trees to fit.                                                                                                                                                       & \{10, 50, 100, 200\}                                                                  \\ \cmidrule{3-5}
                                                                       &                                                                             & num\_leaves                                                       & \begin{tabular}[c]{@{}l@{}}The maximum tree leaves for \\ base learners.\end{tabular}                                                                                                     & \{None, 5, 10\}                                                                       \\ \cmidrule{3-5}
                                                                       &                                                                             & learning\_rate                                                    & \begin{tabular}[c]{@{}l@{}}The Boosting learning rate that \\ controls  how quickly the model \\ adjusts to the  error  in each  \\ iteration of training\end{tabular}                    & \{0.1, 0.01, 0.001\}                                                                  \\ \bottomrule
    \end{tabular}
    }
\end{table}

\smallskip\noindent To evaluate the impact of different classifiers and hyper-parameter tuning on the performance of KUM+TM, we experiment with a diverse set of classifiers:  Naive Bayes (NB), K-Nearest Neighbor (KNN), Decision Trees (DT), Random Forest (RF), XGBoost (XGB), and LightGBM (LGBM). XGB and LGBM represent state-of-the-art classifiers known for their advanced capabilities, while the other classifiers are well-established and widely used in empirical software engineering research.

\begin{figure}[!t]
	\centering
	\includegraphics[width=1.0\linewidth]{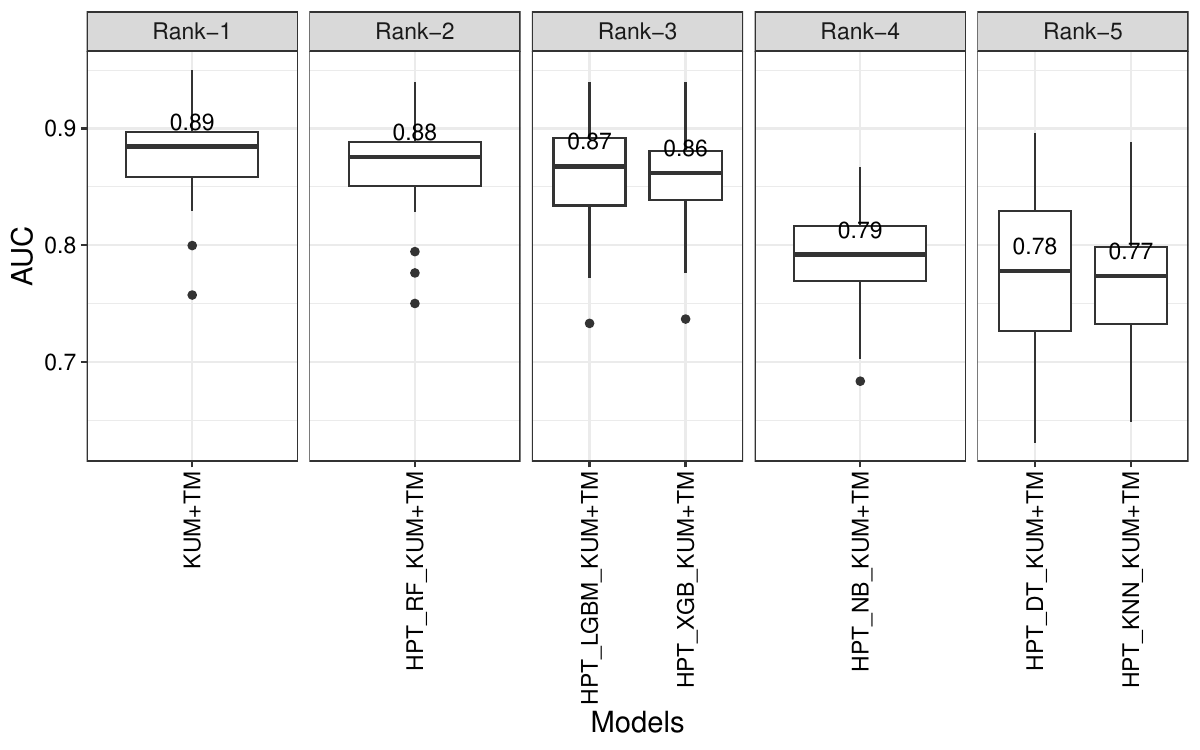}
	\caption{The AUC distribution of hyper-parameter-tuned models and our original
    KUM+TM (random forest with default hyper-parameter values). The names of hyper-parameter tuning models begin with ``HPT\_X'', where X represents
    the name of one of the classifier we studied. The models are grouped based on their performance rankings determined by the Scott-Knott ESD (SK-ESD) method, where a lower SK-ESD rank indicates a better-performing model.} 
	\label{fig:hyper_param_model_sk_rank}
\end{figure}

\sloppy We employ the out-of-sample bootstrap model validation technique with 100 repetitions to ensure robust evaluation. In each iteration, a bootstrap sample is generated to train the model, and the model is tested on the out-of-sample data (i.e., data not included in the bootstrap sample). To identify the optimal configuration for each classifier, we use Scikit-learn's GridSearchCV\footnote{\url{https://scikit-learn.org/stable/modules/generated/sklearn.model_selection.GridSearchCV.html}}. This method performs an exhaustive search over a predefined grid of hyper-parameters (detailed in Table~\ref{tab:hyper_parameter_config}) and employs 10-fold cross-validation to evaluate each configuration. GridSearchCV is applied to the bootstrap sample data, identifying the model configuration that achieves the best performance based on hyper-parameter tuning. The best-tuned model is then tested on the out-of-sample data.

For model evaluation, we follow the methodology described in Step 1.4 of Section~\ref{sec:section_rq1}, ensuring consistency in assessing the effectiveness of the classifiers and their tuned configurations.


Figure~\ref{fig:hyper_param_model_sk_rank} presents the AUC distribution of KUM+TM and the hyper-parameter-tuned variations. The KUM+TM with the default parameter settings is the top ranked one compared to the other hyper-parameter-tuned variations. The hyper-parameter-tuned random forest model (HPT\_RF\_KUM+TM) ranks the second achieving a median AUC of 0.88 which is lower than KUM+TM (the median AUC of KUM+TM is 0.89). In contrast, models built using traditional classifiers such as Naive Bayes (NB), Decision Trees (DT), and K-Nearest Neighbor (KNN) exhibit suboptimal performance, with median AUC scores below 0.80. Of these, the KNN-based model performs the worst, with a median AUC below 0.76. Advanced classifiers like XGBoost and LightGBM achieve better results, with median AUC scores exceeding 0.85, yet they still fail to surpass the performance of the original KUM+TM model. Thus, simply switching to a different classifier with tuned hyper-parameters does not improve the performance of KUM+TM.

\begin{footnotesize}
    \begin{mybox}{Summary}
    	The performance of our original KUM+TM built with the default parameter settings with RF classifier ranks the top performing model among the models built with different classifiers and tuned hyper-parameters. Thus, our original KUM+TM does not improve by simply using a different classification algorithm with tuned hyper-parameters.
    \end{mybox}
\end{footnotesize}

\end{document}